%% file: nirvandels.tex
\documentclass[fleqn,usenatbib]{mnras}
\usepackage{newtxtext,newtxmath}
\usepackage{verbatim}

\usepackage[T1]{fontenc}

\DeclareRobustCommand{\VAN}[3]{#2}
\let\VANthebibliography\thebibliography
\def\thebibliography{\DeclareRobustCommand{\VAN}[3]{##3}\VANthebibliography}

\usepackage{graphicx}	
\usepackage{amsmath}	
\usepackage{subfiles}   
\usepackage{booktabs}   
\usepackage{natbib}     
\usepackage{orcidlink}
\usepackage{tabularx}

\newcommand{\Gyr}{\,{\rm Gyr}}
\newcommand{\Myr}{\,{\rm Myr}}

\newcommand{\hgamma}{\mbox{H\,{\sc $\gamma$}}}
\newcommand{\hbeta}{\mbox{H\,{\sc $\beta$}}}
\newcommand{\halpha}{\mbox{H\,{\sc $\alpha$}}}
\newcommand{\oiii}{\mbox{[O\,{\sc iii]}{$\lambda\lambda 4959,5007$}}}
\newcommand{\oiiia}{\mbox{[O\,{\sc iii]}{$\lambda 5007$}}}

\newcommand{\oii}{\mbox{[O\,{\sc ii]}{$\lambda\lambda 3726,3729$}}}

\newcommand{\neiii}{\mbox{[Ne\,{\sc iii]}{$\lambda 3870$}}}

\newcommand{\oiiinwl}{\mbox{[O\,{\sc iii]}}}
\newcommand{\oiinwl}{\mbox{[O\,{\sc ii]}}}

\newcommand{\neiiinwl}{\mbox{[Ne\,{\sc iii]}}}

\newcommand{\oinwl}{\mbox{[O\,{\sc i]}}}
\newcommand{\niinwl}{\mbox{[N\,{\sc ii]}}}
\newcommand{\siinwl}{\mbox{[S\,{\sc ii]}}}

\newcommand{\zstar}{$Z_{\star}$}
\newcommand{\zgas}{$Z_{\mathrm{g}}$}

\newcommand{\mstar}{$M_{\star}$}

\defcitealias{Weinberg_2017}{W17}
\defcitealias{Bian_2018}{B18}
\defcitealias{Curti_2020}{C20}
\defcitealias{Sanders_2023}{S24}

\title[Stellar and gas-phase metallicities at $z\simeq3.5$]{The NIRVANDELS Survey: the stellar and gas-phase mass-metallicity relations of star-forming galaxies at $\mathbf{z=3.5}$}

\author[T. M. Stanton et al.]{
T. M. Stanton \orcidlink{0000-0002-0827-9769},$^{1}$\thanks{E-mail: t.stanton@ed.ac.uk}
F. Cullen \orcidlink{0000-0002-3736-476X},$^{1}$ 
R. J. McLure,$^{1}$
A. E. Shapley,$^{2}$
K. Z. Arellano-C\'ordova \orcidlink{0000-0002-2644-3518},$^{1}$ \and \ 
R. Begley \orcidlink{0000-0003-0629-8074},$^{1}$  
R. Amor\'in,$^{3}$
L. Barrufet,$^{1}$
A. Calabr\`o,$^{4}$
A. C. Carnall,$^{1}$
M. Cirasuolo,$^{5}$,
J. S. Dunlop,$^{1}$ \and \
C. T. Donnan \orcidlink{0000-0002-7622-0208},$^{1}$ 
M. L. Hamadouche \orcidlink{0000-0001-6763-5551},$^{1}$ 
F. Y. Liu,$^{1}$ 
D. J. McLeod \orcidlink{0000-0003-4368-3326},$^{1}$
L. Pentericci,$^{4}$
L. Pozzetti,$^{6}$ \and \
R. L. Sanders,$^{7}$ 
D. Scholte \orcidlink{0000-0002-6867-1244},$^{1}$ 
and M. W. Topping$^{8}$
\\
$^{1}$Institute for Astronomy, University of Edinburgh, Royal Observatory, Edinburgh, EH9 3HJ, UK\\
$^{2}$Department of Physics \& Astronomy, University of California, 430 Portola Plaza, Los Angeles CA 90095, USA\\
$^{3}$ARAID Foundation. Centro de Estudios de F\'{\i}sica del Cosmos de Arag\'{o}n (CEFCA), Unidad Asociada al CSIC, Plaza San Juan 1, E--44001 Teruel, Spain\\
$^{4}$INAF – Osservatorio Astronomico di Roma, via Frascati 33, 00078, Monteporzio Catone, Italy\\
$^{5}$European Southern Observatory, Karl-Schwarzschild-Str 2, D-86748 Garching b. München, Germany\\
$^{6}$INAF-Osservatorio di Astrofisica e Scienza dello Spazio, Via Gobetti 93/3, I-40129 Bologna, Italy\\
$^{7}$Department of Physics and Astronomy, University of Kentucky, 505 Rose Street, Lexington, KY 40506, USA\\
$^{8}$Steward Observatory, University of Arizona, 933 N Cherry Avenue, Tucson, AZ 85721, USA
}

\date{Accepted XXX. Received YYY; in original form ZZZ}

\pubyear{2024}

\begin{document}
\label{firstpage}
\pagerange{\pageref{firstpage}--\pageref{lastpage}}
\maketitle

\begin{abstract}
We present determinations of the gas-phase and stellar metallicities of a sample of 65 star-forming galaxies at $z \simeq 3.5$ using rest-frame far-ultraviolet (FUV) spectroscopy from the VANDELS survey in combination with follow-up rest-frame optical spectroscopy from VLT/KMOS and Keck/MOSFIRE.
We infer gas-phase oxygen abundances (\zgas; tracing O/H) via strong optical nebular lines and stellar iron abundances (\zstar; tracing Fe/H) from full spectral fitting to the FUV continuum.
Our sample spans the stellar mass range $8.5 < \mathrm{log}(M_{\star}/\mathrm{M}_{\odot}) < 10.5$ and shows clear evidence for both a stellar and gas-phase mass-metallicity relation (MZR).
We find that our O and Fe abundance estimates both exhibit a similar mass-dependence, such that $\mathrm{Fe/H}\propto M_{\star}^{0.30\pm0.11}$ and $\mathrm{O/H}\propto M_{\star}^{0.32\pm0.09}$.
At fixed $M_{\star}$ we find that, relative to their solar values, O abundances are systematically larger than Fe abundances (i.e., $\alpha$-enhancement).
We estimate an average enhancement of $\mathrm{(O/Fe)} = 2.65 \pm 0.16 \times \mathrm{(O/Fe)_\odot}$ which appears to be independent of $M_{\star}$.
We employ analytic chemical evolution models to place a constraint on the strength of galactic-level outflows via the mass-outflow factor ($\eta$).
We show that outflow efficiencies that scale as $\eta \propto M_{\star}^{-0.32}$ can simultaneously explain the functional form of of the stellar and gas-phase MZR, as well as the degree of $\alpha$-enhancement at fixed Fe/H.
Our results add further evidence to support a picture in which $\alpha$-enhanced abundance ratios are ubiquitous in high-redshift star-forming galaxies, as expected for young systems whose interstellar medium is primarily enriched by core-collapse supernovae.
\end{abstract}

\begin{keywords}
galaxies: abundances -- galaxies: evolution -- galaxies: high redshift 
\end{keywords}



\section{Introduction}
\label{sec:intro}

\subfile{sections/1-Introduction.tex}

\section{Data and Sample Properties}
\label{sec:data}

\subfile{sections/2-Data.tex}

\section{Determining gas-phase and stellar metallicities} 
\label{sec:analysis}

\subfile{sections/3-Analysis.tex}

\section{The stellar and gas-phase mass-metallicity relations}
\label{sec:results}

\subfile{sections/4-MZRs.tex}

\section{Constraining the mass-dependence of galaxy outflows}
\label{sec:gce-modelling}

\subfile{sections/5-Outflows.tex}

\section{Conclusions}
\label{sec:conclusions}

\subfile{sections/6-Conclusions.tex}

\section*{Acknowledgements}

    T. M. Stanton, F. Cullen, K. Z. Arellano-C\'ordova and D. Scholte acknowledge support from a UKRI Frontier Research Guarantee Grant (PI Cullen; grant reference: EP/X021025/1). 
    R. J. McLure, R. Begley, L. Barrufet, J. S. Dunlop, C. T. Donnan, M. L. Hamadouche, F. Y. Liu and D. J. McLeod acknowledge the support of the Science and Technology Facilities Council. 
    J. S. Dunlop also thanks the Royal Society for their support through a Royal Society Research Professorship. 
    A. C. Carnall thanks the Leverhulme Trust for their support via the Leverhulme Early Career Fellowship scheme.
    \par This research utilised \textsc{astropy}, a community-developed core Python package for Astronomy \citep{astropy}, \textsc{NumPy} \citep{Numpy}, \textsc{SciPy} \citep{Scipy}, \textsc{Matplotlib} \citep{Hunter:2007}, \textsc{dynesty} \citep{dynesty_Speagle} and NASA's Astrophysics Data System Bibliographic Services.
    
    For the purpose of open access, the author has applied a Creative Commons Attribution (CC BY) licence to any Author Accepted Manuscript version arising from this submission.

\section*{Data Availability}

    The VANDELS spectroscopic data utilised in this work is publicly available, and can be accessed through the \href{http://archive.eso.org/scienceportal/home}{ESO Science Portal}. All other data will be shared by the corresponding author upon reasonable request.
    

\bibliographystyle{mnras}
\bibliography{nirvandels.bib} 


\appendix

    \section{Systematic uncertainties in determining galaxy abundances.}\label{app:systematic_uncertainties}
        
        \begin{figure*}
                \centering
                \includegraphics[width=\linewidth]{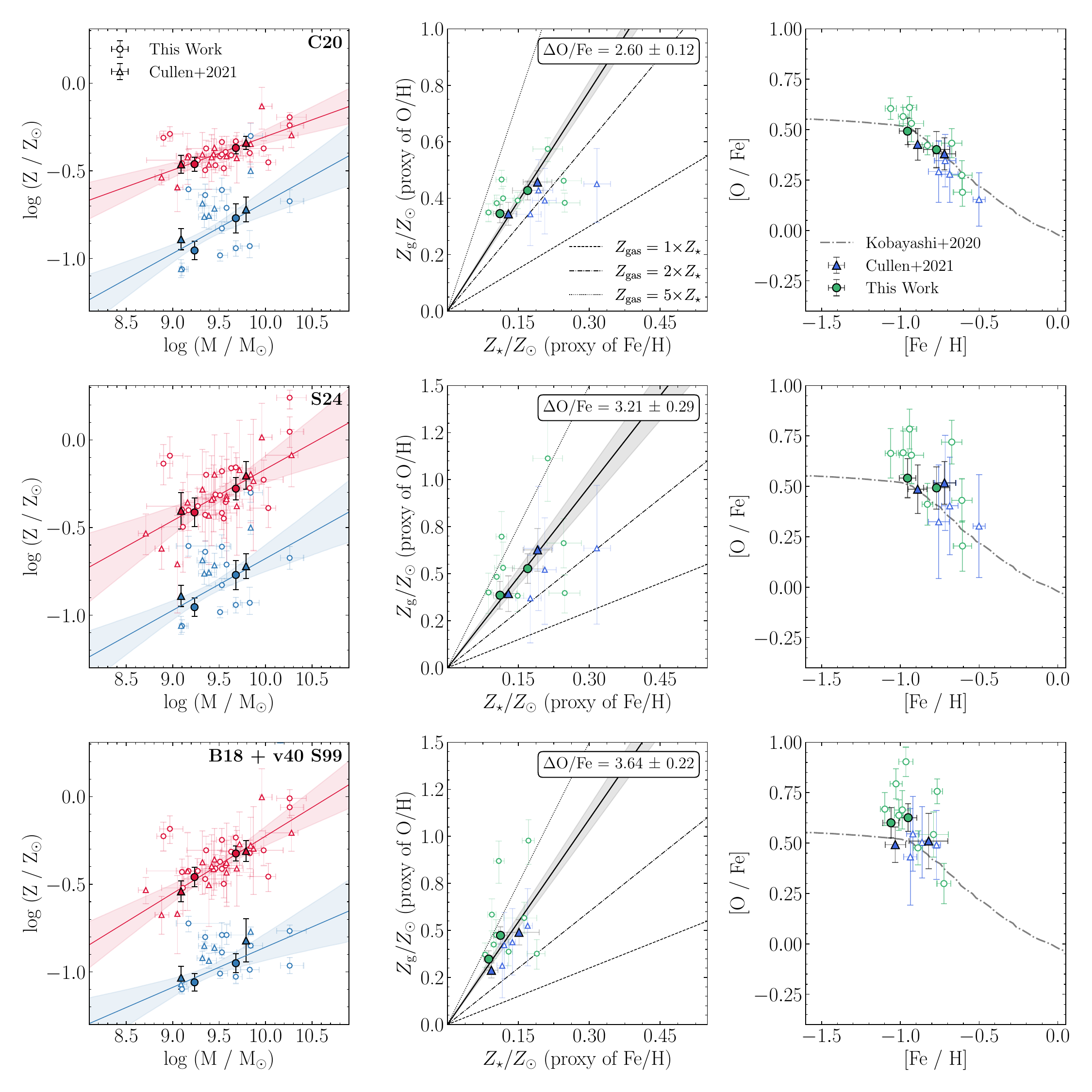}
                \caption{Comparison of the effects of different strong-line calibration scheme and stellar population model choices on the on the key results of this work.
                The first row corresponds to the \citetalias{Curti_2020} calibration scheme, and the second corresponds to the \citetalias{Sanders_2023} calibration scheme. The third row uses the \citetalias{Bian_2018} calibrations, but substitutes the no-rotation S99 models for the v40-rotation S99 models. 
                The left column emulates Fig.~\ref{fig:combined-mzr}, and the central and right columns emulate the left and right panels of Fig.~\ref{fig:zgas-vs-zstar} respectively. 
                The \citetalias{Curti_2020} calibration finds a shallower MZ$_\mathrm{g}$R and hence a lower degree of alpha enhancement, whilst the \citetalias{Sanders_2023} calibrations yield a much steeper MZ$_\mathrm{g}$R and hence a high degree of enhancement. Switching to rotational stellar population models reduces the normalisation of the stellar MZR, increasing the degree of alpha enhancement. All three configurations maintain the overlap of our high-z systems with the knee of the \citet{Kobayashi_2020} relation.
                }
                \label{fig:zg-zs-comparison-s99}
        \end{figure*}  

    \subfile{tables/app1.tex}

    The primary results of this work utilised the \citetalias{Bian_2018} calibration scheme and Starburst99 non-rotating models (v00).
    Here, we show how different choices can affect the primary result of this work.
    In Fig. \ref{fig:zg-zs-comparison-s99} we show versions of our main results for the following combinations of strong-line calibration and stellar population model: (i) the \citetalias{Curti_2020} calibrations (derived from local SDSS star-forming galaxies, to which we have added the \citet{Curti_2023} Ne3O2 ratio calibration) with the Starbust99-v00 models; (ii) the \citetalias{Sanders_2023} calibration (defined using auroral line measurements from early JWST measurements of galaxies at $2 \leq z \leq 9$) with the Starbust99-v00 models and, (iii) our fiducial \citet{Bian_2018} calibration and Starburst99-v40 (i.e. rotating) models. 
    The abundances corresponding to these systematic choices for the individual and composite samples can be found in Table~\ref{tab:alternative-abundances}.

    We find that slope of the stellar MZR is the same for both the rotating and non-rotating models, but that the stellar Fe abundances of the rotating models are $\simeq 0.1$ dex lower.
    For the O abundances derived from strong line calibrations, we find that all three calibrations yield a consistent median value of ${\mathrm{dlog}(Z_{\mathrm{g}}/\mathrm{Z}_{\odot}) \simeq 0.4}$ but that the slope and scatter of the MZR$_{\mathrm{g}}$ are clearly affected.
    We find that the \citet{Curti_2020} calibrations yield the smallest scatter and shallowest slope with ${\mathrm{d(log}Z_{\mathrm{g}})/dM_{\star}=0.16 \pm 0.07}$.
    Ultimately, all combinations predict ubiquitous $\alpha$-enhancement with average values int he range $\langle \mathrm{O/Fe} \rangle \simeq 2-3 \times \mathrm{(O/Fe)_\odot}$.
    However, the shallow slope given by the \citet{Curti_2020} calibrations implies a decrease in $\mathrm{O/Fe}$ at higher stellar masses, which would have important consequences for interpreting the outflow strengths and star-formation timescales.

    In the near future \emph{JWST} programs such as AURORA (PI Shapley, Sanders; GO1914) and EXCELS (PI Carnall, Cullen; GO3543) will help to resolve this issue by combining direct T$_{\mathrm{e}}$-based O/H and stellar Fe/H estimates for samples of galaxies at $z>2$.
    

\bsp	
\label{lastpage}
\end{document}

%% file: sections/1-Introduction.tex
    The secular processes that govern the growth of galaxies are dictated by an overarching cosmic baryon cycle, which encompasses the infall of gas into galaxies, its collapse into and processing by stars, and its release via outflows driven by stellar and active galactic nuclei (AGN) feedback.
    The metal enrichment of gas and stars in galaxies is intrinsically linked to each of these processes, and therefore the combination of galaxy metallicities with measurements of galaxy properties, such as stellar mass, provides a powerful tool for studying the evolution and growth of galaxies \citep{Maiolino_Mannucci_2018, Peroux_2020}.

    \par

    The gas-phase abundance of oxygen relative to hydrogen (O/H) is the most common proxy for galaxy metallicity because oxygen dominates the elemental mass distribution and can be readily constrained from rest-frame optical spectra \citep{Kewley_2019}. 
    At high redshifts, oxygen abundances are usually inferred from nebular emission line ratios that have been calibrated against direct $T_e$ method abundances \citep[e.g.,][]{Maiolino_Mannucci_2018}.
    There are a variety of these calibrations, each of which is sensitive to different metallicity regimes and subject to their own systematic uncertainties \citep[e.g.,][]{Curti_2017, Bian_2018, Sanders_2023}.
    Even in the local Universe these systematics are large \citep[up to 0.7 dex in log(O/H); e.g.,][]{Kewley_2008, Arellano_Cordova_2022}, and the situation is compounded at high redshift where significant samples of direct abundances (needed to derive robust empirical calibrations) are only now becoming available with \emph{JWST} observations \citep[e.g.,][]{Sanders_2023}.

    However, relative abundances derived from a given calibration at a given redshift are presumed to be reliable \citep{Kewley_2019} and, across the literature, gas-phase metallicities (\zgas) have been measured from the local universe out to redshifts $z\sim10$ \citep[e.g.,][]{Curti_2023}.
    At all redshifts, a tight correlation between gas-phase oxygen abundance and galaxy stellar mass (\mstar) has been confirmed (i.e., the mass-metallicity relationship, MZR) such that galaxies at higher masses exhibit higher gas-phase metallicities \citep[e.g.,][]{Tremonti_2004, Andrews_2013, Curti_2020}. 
    Furthermore, multiple studies have shown that the MZR evolves with redshift, such that at fixed stellar mass galaxies at higher redshifts are less oxygen enriched \citep[e.g.,][]{Savalgio_2005, Cullen_2014, Troncoso_2014, Sanders_2021, Li_2023}.
    While the existence of the MZR is robustly established, fully characterising and understanding the shape and evolution of the MZR with cosmic time remains a key goal in observational astrophysics.

    \par

    An alternative tracer of the metal content of high-redshift galaxies is the abundance of iron (Fe/H), which can be derived from photospheric line blanketing in the rest-frame far ultraviolet (FUV) continuum \citep[e.g.,][]{Leitherer_2010, Steidel_2016, Cullen_2019}.
    The rest-frame FUV traces young, massive, O- and B- type stars, and FUV-based stellar metallicities reflect the Fe/H of the interstellar medium (ISM) from which they formed, permitting a direct comparison between the Fe/H of these stars and the gas-phase O/H derived the rest-frame optical nebular spectra.
    \citet{Steidel_2016} introduced a novel method for deriving stellar Fe/H from full spectra fitting to rest-frame FUV spectra, and this method has subsequently been developed to robustly determine stellar metallicities up to $z \simeq 3.5$ \citep[e.g.,][]{Cullen_2019}.
    A number of independent studies have now confirmed the existence of the stellar MZR at $z \simeq 2 - 3.5$ \citep[e.g.,][]{Cullen_2019, Calabrò_2021, Kashino_2022, Chartab_2023} with a shape similar to the local relation, but showing an evolution to lower metallicity \citep[e.g.,][]{Galazzi_2005, Panter_2008, Zahid_2017}.

    \par

    Combining rest-frame FUV and rest-frame optical spectroscopy with the aforementioned methodologies, it is possible to simultaneously determine both O/H and Fe/H in star-forming galaxies, yielding an estimate of the O/Fe abundance ratio \citep[e.g.,][]{Steidel_2016, Topping_2020A, Topping_2020B, Cullen_2021}.
    The O/Fe ratio is sensitive to star-formation timescales due to the fact that oxygen (and $\alpha$-elements in general) are produced primarily by core-collapse supernovae (CCSNe) over short timescales, whereas Fe peak elements are released by \emph{both} CCSNe and Type-Ia supernovae (SNIa) probing longer timescales \citep[e.g.,][]{Maoz_Mannucci_2012, Kobayashi_2020}. 
    Determining O/Fe therefore provides insight on the relative contributions of CCSNe and SNIa, such that galaxies exhibiting super-solar O/Fe ratios (often referred to more generally as $\alpha$-enhancement) display chemical enrichment patterns dominated by the yields of CCSNe.
    As such, a number of authors have highlighted a connection between O/Fe ratios and the specific star-formation rate of galaxies \citep[e.g.][]{matthee2018, Kashino_2022, Chruslinska_2023}.
    Additionally, chemical evolution models predict that the elemental abundance ratios are sensitive to various macroscopic galaxy processes, such as outflow and star-formation efficiencies \citep[e.g.,][]{Weinberg_2017}.
    As a result, the combination of absolute (e.g., Fe/H) and relative (e.g., O/Fe) abundances can be used to explore variations in outflow properties and star-formation histories across the galaxy population. 
    
    \par

    Moreover, as discussed in \citet{Topping_2020A}, an Fe deficit relative to O also offers a natural explanation for the evolution in ISM conditions at high redshift, as indicated, for example, by the evolution in classic line ratio diagnostic diagrams (e.g., the \niinwl \ BPT, \citealp{Steidel_2014, Shapley_2019}; the \siinwl \ BPT, \citealp{Shapley_2019}; \neiiinwl \ emission-line diagrams, \citealp{Jeong_2020}; and the \oinwl \ BPT, \citealp{Clarke_2023}).
    For $\alpha$-enhanced galaxies, the stellar ionising spectrum at fixed O abundance is harder, as lower iron abundance leads to less metal-line blanketing in the ionising continuum at $< 912${\AA}.  
    Accurate nebular modelling of galaxies at high redshift therefore requires robust estimates of the typical O/Fe ratio as a function of galaxy properties.

    \par 

    The number of studies exploring O/Fe ratios in high redshift galaxies (${z \simeq 2-4}$) has continued to grow since the pioneering work of \citet{Steidel_2016}.
    To date, all works find ubiquitous evidence for $\alpha$-enhanced abundance ratios in the range $\simeq 2 - 5 \times \mathrm{(O/Fe)_\odot}$ either via estimates of O and Fe abundances for the same sources \citep[e.g.,][]{Topping_2020A, Topping_2020B, Cullen_2021, Chartab_2023} or indirect approaches \citep[such as photoionisation modeling or comparison of different sources;][]{Sanders_2020, Strom_2022, Kashino_2022}.
    These direct galaxy estimates of $\alpha$-enhancement are in good agreement with the signal measured in high-redshift Damped Ly$\alpha$ abosrption line systems \citep[e.g.,][]{cooke2011, de_cia2016, Velichko_2024}.
    These results corroborate expectations that young star-forming galaxies at high redshift are $\alpha$-enhanced and reflect the dominance of CCSNe in star-forming galaxies at these epochs \citep{Maiolino_Mannucci_2018}.
    Interestingly, the estimated level of $\alpha-$enhancement at these redshifts has been shown to be consistent with stellar archaeological measurements of $10-12$ Gyr old stellar populations in the Milky Way \citep{Cullen_2021}.

    \par 
    However, despite this progress, the number of individual galaxies with O/Fe estimates is still relatively small \citep[e.g.,][]{Topping_2020A, Topping_2020B}, and explorations of the variation of $\alpha$-enhancement with global galaxy properties have been limited.
    In this work, we compile a sample of $65$ star-forming galaxies at $3.0 < z < 3.8$ with simultaneous deep rest-frame optical spectroscopy (from VLT/KMOS and Keck/MOSFIRE) and FUV spectroscopy \citep[from the VANDELS survey;][]{McLure_2018, Pentericci_2018, Garilli_2021} to determine O and Fe abundances for both individual and composite spectra.
    Our new analysis follows on from \citet{Cullen_2021} who used a sample of $33$ galaxies at $z\simeq3.5$ to determine an average degree of $\alpha$-enhancement of $\mathrm{(O/Fe)}\simeq 2.5 \times \mathrm{(O/Fe)_\odot}$.
    In this paper, we double the sample size and expand on the analysis in \citet{Cullen_2021} by improving constraints on the gas-phase and stellar MZR at $z=3.5$, as well as the variation of O/Fe with stellar mass.
    We also demonstrate how our combination of absolute and relative abundance determinations enables us to place novel constraints on the mass scaling of large-scale galactic outflows via comparison with analytic chemical evolution models.
    
    \par

    The structure of this paper is as follows. 
    In Section~\ref{sec:data}, we discuss our sample selection and the primary spectroscopic datasets used in this work.
    Section~\ref{sec:analysis} describes the methodologies that we have employed to determine the abundances of O and Fe from the spectra. 
    In Section~\ref{sec:results} we present our determinations of the gas-phase and stellar MZRs at $z \simeq 3.5$, along with our inferred measurements of the O/Fe ratios for our sample.
    We then present constraints on the stellar mass dependence of galactic-scale outflows using analytical chemical evolution models.
    Our conclusions are summarised in Section~\ref{sec:conclusions}.

    \par

    Throughout this work, metallicities are quoted relative to a solar abundance value taken from \citet{Asplund_2009}, which has a bulk composition by mass of $Z_* = 0.0142$ and relates the widely-used 12~+~log(O/H) scale to the \zgas \ scale via the expression $\mathrm{12 + log(O/H)}$ = log({\zgas}) + 8.69. 
    We assume the following cosmology: $\Omega_M = 0.3$, $\Omega_\Lambda = 0.7$, and $H_0 = 70 \ \mathrm{km s}^{-1} \mathrm{Mpc}^{-1}$.

%% file: sections/2-Data.tex
        \subfile{../tables/table1.tex}

        Our analysis is based on Keck/MOSFIRE and VLT/KMOS near-IR spectroscopic follow-up of galaxies selected from the VANDELS survey \citep{McLure_2018}.
        The VANDELS spectroscopy (taken with the VLT/VIMOS spectrograph) provides rest-frame FUV spectra for each galaxy while the near-IR spectroscopy traces the rest-frame optical.
        For $\simeq 50$ per cent of our final galaxy sample we obtained near-IR follow-up with Keck/MOSFIRE and these observations have been previously described in detail in \citet{Cullen_2021}.
        The new data presented here consist of 40 hours of VLT/KMOS observations (ID:108.21Z5.001; PI F. Cullen) obtained between October 2021 and February 2023.
        In the following, we describe each of these datasets in turn, with a particular focus on the selection and reduction of the new KMOS observations used in this paper.

        \begin{figure*}
            \vspace{-0.3cm}
            \centering
            \includegraphics[width=\linewidth]{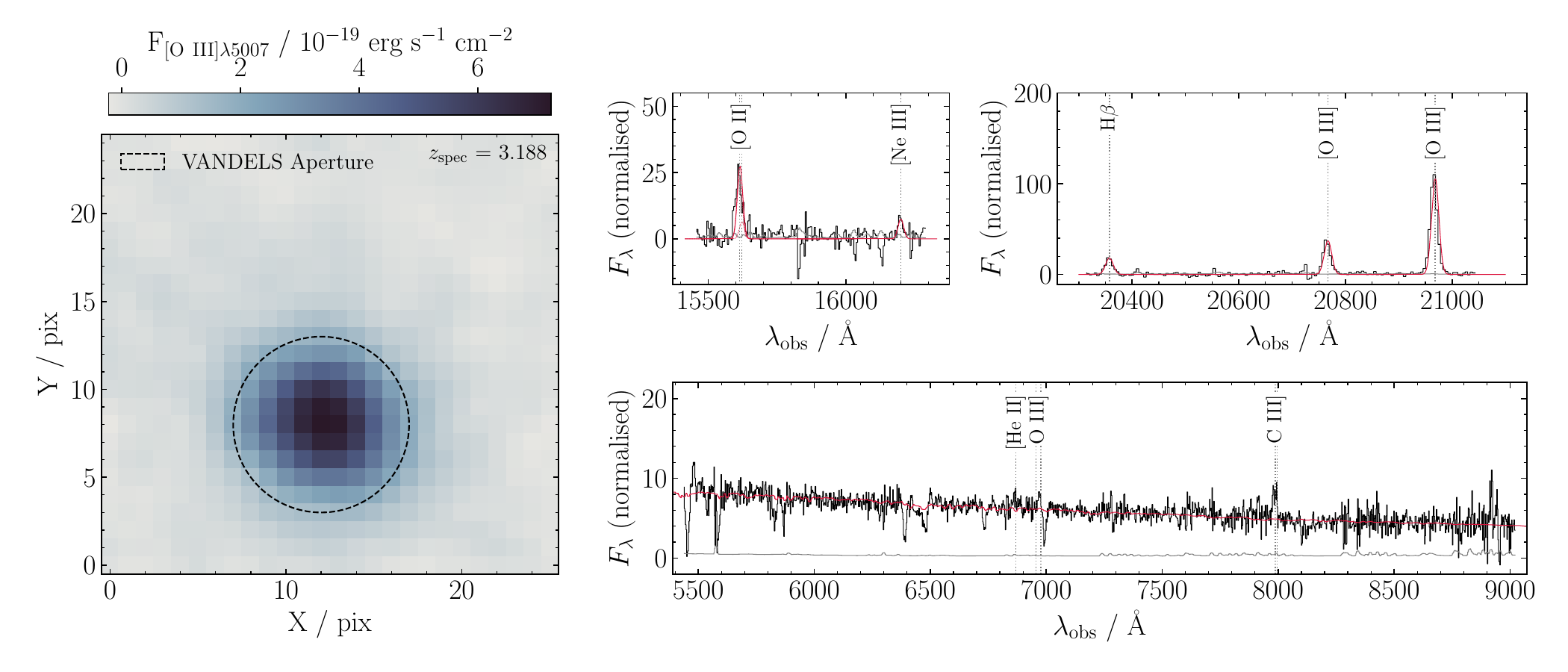}
            \caption{An example galaxy from our VANDELS plus VLT/KMOS sample. 
            The left-hand panels shows a 2D slice through the KMOS datacube centred on the wavelength of the \oiiia \ emission line for the galaxy KVS-208  ($z=3.188$; Table \ref{tab:bian.s99}).
            The dashed circle represents the 1 arcsecond diameter aperture used to extract the galaxy integrated spectra selected to match to the extraction for the VANDELS spectra.
            The top two panels in the upper right-hand side show two regions from the extracted KMOS spectrum centred on the \oii, \neiii, \hbeta \ and \oiii \ emission lines used to derived gas-phase O/H (see Section~\ref{sec:determine_zgas}).
            The black line shows the extracted spectrum and the red line shows the best-fitting model assuming Gaussian line profiles and a flat continuum (see Section \ref{sec:dqs} for details of the emission-line fitting).
            The lower right-hand panel shows the rest-frame FUV VANDELS spectrum for this object.
            The VANDELS spectrum is shown in black and the associated error spectrum in grey.
            The positions of nebular FUV emission lines are marked by vertical dashed lines.
            The best fitting Starburst99 stellar population model from which we estimate the stellar Fe/H is shown in red (see Section~\ref{sec:determine_zstar} for details of the FUV continuum fitting).}
            \label{fig:example-kmos}
        \end{figure*}

        \subsection{The VANDELS sample}

        Star-forming galaxies at $z\simeq3.5$, observed as part of the VANDELS survey, form the basis of our analysis.
        VANDELS was an ESO/VLT public spectroscopic survey conducted with the VIMOS spectrograph \citep{Pentericci_2018, McLure_2018, Garilli_2021}.
        VANDELS targets were selected from regions centred on the CANDELS CDFS and UDS fields \citep{Grogin_2011, Koekemoer_2011} in three categories: (i) massive passive galaxies at $1.0 \leq z \leq 2.5$, (ii) bright star-forming galaxies at $2.4 \leq z \leq 5.5$ and (iii) fainter star-forming galaxies at $3.0 \leq z \leq 7.0$, with the primary focus being star-forming galaxies at $z>2.4$ (comprising $85\%$ of all targets).
        All observations were obtained using the VIMOS medium resolution grism covering the wavelength range $0.48 \mu \mathrm{m} < \lambda_{\mathrm{obs}}  < 1.0 \mu \mathrm{m}$ at a median resolution of $\mathrm{R}=580$ (with 1 arcsec slits) and a dispersion of $2.5${\AA} per pixel.
        The median seeing across all observations was $0.7$ arcsec.

        In our redshift range of interest ($3.0 < z < 3.8$) the VANDELS spectra cover the rest-frame wavelength range $1000$ \AA \ $< \lambda < 2000$ {\AA}.
        The depth of the VANDELS observations (median 40hr integration per galaxy) enables stellar Fe/H to be derived for both individual and composite spectra of galaxies \citep[e.g.,][]{Cullen_2019, Calabrò_2021, Cullen_2021} as we describe in Section \ref{sec:determine_zstar}.
        Full descriptions of the VANDELS survey design and target selection can be found in \citet{McLure_2018}, and observations and data reduction are described \citet{Pentericci_2018} and \citet{Garilli_2021}.

        \subsection{MOSFIRE Observations}

        For $33$ galaxies in our final sample, we obtained rest-frame near-IR spectroscopic follow-up with the Multi-Object Spectrometer for Infrared Exploration \citep[MOSFIRE;][]{McLean_2012} on the Keck~I telescope.
        We observed the VANDELS-selected targets in the redshift range $3.0 < z < 3.8$ with the Keck/MOSFIRE $H$ and $K$ bands in order to target the \oii, \neiii, \hbeta \ and \oiiia \ nebular emission lines necessary to derive O/H estimates in the ionised gas.
        Observations were conducted using a slit width of $0.7$ arcsec, yielding a spectral resolution of $3650$ in $H$ and $3600$ in $K$.
        The median seeing across all observations was $0.5$ arcsec.
        Despite the narrower MOSFIRE slits compared to VIMOS, the improved seeing for the MOSFIRE observations means that both observations probe comparable intrinsic regions in our target galaxies.
        Full details of the selection and properties of the Keck/MOSFIRE sample are given in \citet{Cullen_2021}.

        \subsection{KMOS Observations}
        \label{sec:vandels_ob}

            The new observations presented here consist of additional near-IR follow-up of VANDELS targets at $3.0 < z < 3.8$ obtained with the $K$-band Mult-Object Spectrograph (KMOS) on the VLT.
            Below we describe the sample selection, observations, and data reduction for this sample.

        \subsubsection{Sample Selection}
            
            The initial targets for VLT/KMOS observations were drawn from the VANDELS DR3 catalogue and selected to have: (i) a redshift in the range $3.0 \leq z \leq 3.8$ (so that the \oii, \neiii, \hbeta \ and \oiii \ optical nebular emission lines were accessible to the KMOS $HK$ grating) and (ii) a redshift quality flag $z_{\mathrm{flag}}\geq 2$ (e.g., a redshift flag 2, 3, 4, 9, or 14,  as defined in \citealp{Pentericci_2018}, corresponding to a \ $\geq70$ per cent probability of the redshift being correct).
            From this sample, the highest priority was given to targets according to following criteria: (i) a signal-to-noise ratio (SNR) per resolution element in the VANDELS spectrum of $\geq 3$ (to maximise the number of targets for which it would be possible to measure individual Fe/H); (ii) a redshift quality flag of $z_{\mathrm{flag}}=3$ or $4$ (corresponding to a $\geq95\%$ probability of the redshift being correct; \citealp{Pentericci_2018}); (iii) a redshift such that the \oii, \hbeta \ and \oiii \ optical nebular emission lines were predicted to fall on relatively sky-free regions of the KMOS $HK$ grating and (iv) an estimated stellar mass of $\geq 10^{9}$M$_{\odot}$ (to ensure optimal SNR of the optical lines; see Section \ref{sec:data-mass-sfr} for details on the stellar mass estimates).
            These criteria were imposed to maximise the number of individual galaxies for which Fe/H and O/H could be estimated.
        
        \subsubsection{Observations and Data Reduction}
        \label{sec:kmos_ob}

            KMOS is a near-IR spectrograph consisting of 24 integral field units (IFUs) operating simultaneously within a 7.2 arcmin circular field-of-view (FoV). 
            Each IFU has a $2.8$ arcsec $\times$ $2.8$ arcsec FoV and a uniform spatial sampling of $0.2$ arcsec $\times$ $0.2$ arcsec.
            The $24$ IFUs are fed into $3$ separate detectors with $8$ IFUs assigned per detector.
            Across our multiple observing runs (see Table~\ref{tab:observations}) one IFU was continuously out of commission meaning that we only observed $23$ sources per pointing.
            We observed in the KMOS $HK$-band covering the wavelength range $1.48\,\mu \mathrm{m} - 2.44 \,\mu \mathrm{m}$ with a spectral resolution of $\mathrm{R}=1985$ at band centre.
            A total of two pointings were observed, with $20$ IFUs per pointing assigned to VANDELS-selected galaxies (described above) and $3$ IFUs per pointing assigned to reference stars (one reference star per detector). 
            These reference stars were used to correct for small detector-dependent shifts in the position of targets between exposures (as described below). 

            For all objects we adopted an object-sky-object nodding strategy with $300\mathrm{s}$ per exposure (or observing block, OB) and dithered the exposures for sky sampling and subtraction.
            The resulting maximum on-source integration time per object was $10$ hours.
            Across the full observing run, various IFUs periodically malfunctioned and were out of commission on a given night, resulting in a nonuniform observing time across all objects in the sample.
            We also removed observations taken during poor seeing conditions ($> 0.8$ arcsec) or objects for which the shifts between OBs could not be accurately calibrated using the reference star.
            In practice, $94$ per cent of the OBs in pointing 1 were usable.
            However, due to significant disruption caused by out-of-commission pickoff arms, only $40$ per cent of OBs in pointing 2 were used.
            Details of our VLT/KMOS observations are given in Table~\ref{tab:observations}.
    
            \par
            We reduced the raw data to produce three-dimensional science and error spectra using the \textsc{esorex-kmos} pipeline (v 4.0.4) described in \citet{Davies_2013}. 
            For a given IFU, the pipeline extracts the raw image and performs dark and flatfield corrections, illumination corrections, and wavelength calibrations.
            The pipeline then combines all the reduced images corresponding to the given IFU to form a final datacube.
            To further improve the quality of the data reduction, we implemented two routines in addition to the standard pipeline routines. 
            First, each observing block (OB) was combined using a custom algorithm that accounts for small shifts in the positions of targets between each exposure.
            These small shifts (on average $\lesssim 3$ pixels for the usable OBs) are detector-dependent and were calculated using the position of the bright references stars in each pointing.
            Second, to derive a more accurate estimate of the sky background, we masked the positions of the nebular emission lines in the datacubes (following \citealp{Stott_2016}) and subtracted the combined median sky cube from all of our final datacubes.
            The generation of the final datacubes was run in two passes: first without sky subtraction to produce initial datacubes from which accurate KMOS spectroscopic redshifts could be measured, and then with sky subtraction using the measured redshifts.
            Finally, the data cubes were reconstructed on a spatial scale of $0.1$ arcsec $\times$ $0.1$ arcsec.
                        
            \par
            
            For the analysis in this paper, we focus on galaxy-integrated KMOS spectra.
            To generate the galaxy-integrated one-dimensional spectra, we summed the flux within $1$ arcsec diameter apertures centred on the peak of the \oiiia \ emission.
            An aperture size of $1$ arcsec was found to be a good compromise between sampling as much of the galaxy as possible while also maximising the signal-to-noise in the extracted spectra \citep[see also][]{hayden-pawson2022}.
            It is worth noting that the chosen aperture size should not introduce strong biases in the derived metallicities since galaxies at $z>3$ are expected to have flat metallicity gradients \citep[e.g.,][]{Curti_2020, Venturi_2024}. 
            Crucially, a $1$ arcsec aperture size matches the VANDELS aperture size, ensuring that we are probing the same regions of our galaxies in the rest-frame optical as in the rest-frame FUV.
            An example of a slice through one datacube at the position of the \oiiia \ line with the associated extraction aperture is shown in Fig.~\ref{fig:example-kmos}. 
            We also show the resulting 1D spectra in the region of the \oii, \neiii, \hbeta \ and \oiii \ emission lines.

            \par
            
            We evaluated the accuracy of the flux calibration of the final 1D spectra by comparing the integrated flux of the reference stars to photometry from the \textsc{3D-HST} catalogues \citep{skelton_2014, momcheva_2016} in matched apertures.
            We found good agreement with offsets of $< 10$ per cent averaged across the six reference stars.

    \subsection{Final Galaxy Sample}
    \label{sec:finsamp}

        \subfile{../tables/table2.tex}

        The final VLT/KMOS sample analysed in this paper consists of $32/40$ of the targeted galaxies covering the redshift range $3.0 < z < 3.7$.
        We excluded $7/40$ galaxies that lacked a clear ($> 3 \sigma$) emission line detection required for accurately determining a redshift.
        An additional galaxy was excluded due to lacking spectral coverage of the \oii \ feature required for our abundance determinations.
        All galaxy spectra were visually inspected to exclude obvious AGN. 
        We found no galaxies that exhibited extremely broad emission features or high ionization emission features.
        We also ruled out the presence of significant AGN ionization based on their mid- IR SED shapes and X-ray properties \citep{McLure_2018}.
        An example galaxy from the sample (KVS-208) is shown in Figure~\ref{fig:example-kmos}, and the general properties of the KMOS sample are summarised in Table~\ref{tab:bian.s99}. 

        In addition to the new VLT/KMOS sample, we also include the $33$ galaxies from the Keck/MOSFIRE sample described in \citet{Cullen_2021}.
        As described below, the methodology for deriving key physical parameters (e.g., stellar mass, O/H, Fe/H) is the same between the two samples and therefore, unless explicitly stated, we take estimates of these parameters directly from \citet{Cullen_2021} (see their Table 2).
        
        Combined, our full sample consists of 65 galaxies in the redshift range $2.95 < z < 3.80$ (median $z=3.5$).
        Each galaxy has both rest-frame optical and rest-frame FUV spectra, from which we infer O/H and Fe/H respectively (see Section \ref{sec:analysis}).
        In the instance of being unable to infer either quantity for a given galaxy, the known spectroscopic redshift permits inclusion into the generation of composite spectra (see Section~\ref{sec:compspec}).
        All galaxies are selected from the VANDELS spectroscopic survey.
        As described in a number of previous studies, the star-forming galaxies in the VANDELS survey are consistent with typical `main-sequence' galaxies \citep[e.g.,][]{McLure_2018, Cullen_2021, Garilli_2021} and we have verified that this is also case for the combined sample presented here.
        Our final sample can therefore be considered representative of normal star-forming galaxies at $z \simeq 3.5$ within the mass range of our sample (see Section~\ref{sec:dqs}).

        \begin{figure}
            \vspace{-0.3cm}
            \centering
            \includegraphics[width=\linewidth]{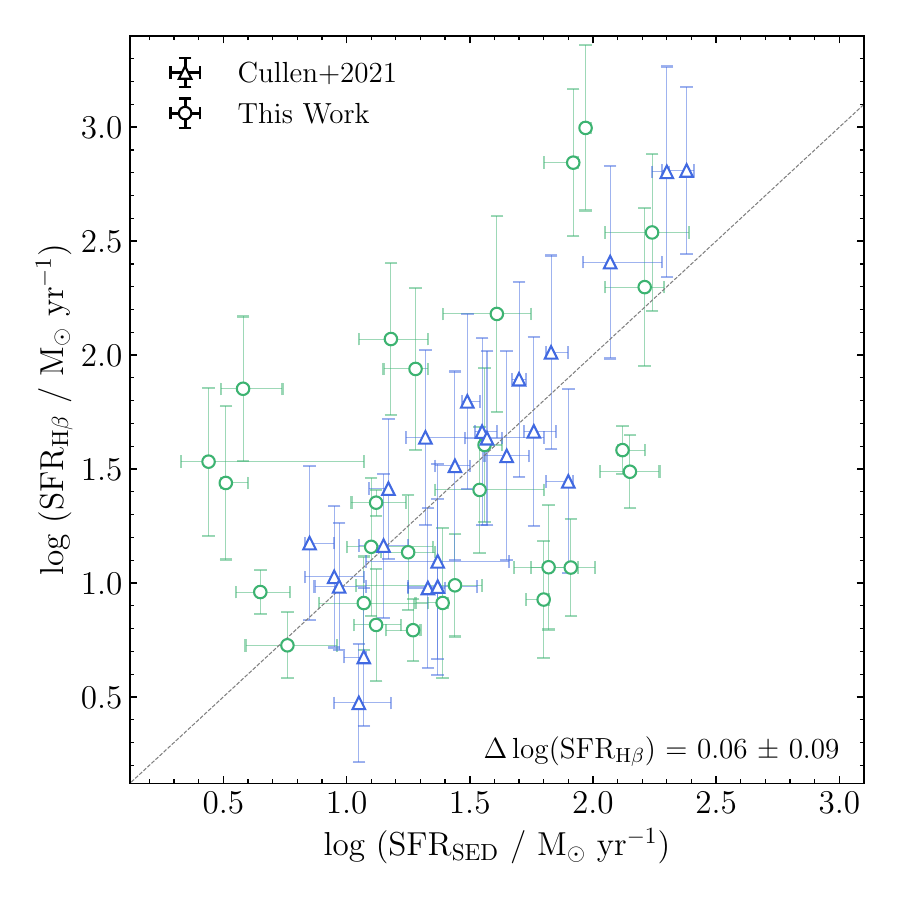}
            \caption{A comparison between the star-formation ratess derived using the emission-line corrected photometry and those derived from dust-corrected \hbeta \ line fluxes using the \citet{Hao_2011} SFR conversion. 
            The dotted black line represents the 1:1 relation. 
            Galaxies from our VLT/KMOS sample (green circles) are in good agreement with the trend seen for the Keck/MOSFIRE sample presented in \citet{Cullen_2021} (blue triangles). 
            Overall, we see excellent agreement between our $\mathrm{SFR_{SED}}$ and SFR$_\mathrm{\hbeta}$ estimates.
            We find an average offset between our SED derived SFRs and our \hbeta \ derived SFRs of $0.06 \pm 0.09$ dex from the 1:1 line indicating that our stellar and nebular dust correction are fully self-consistent.}
            \label{fig:hb-sfr-vs-phot-sfr}
        \end{figure}

    \subsection{Measurements and derived quantities}
    \label{sec:dqs}

        In the following, we describe our measurements of global galaxy properties (e.g., stellar mass, star-formation rate) and emission line fluxes for the galaxies in our VLT/KMOS sample.
        The methodology adopted here closely follows the methodology described in \citet{Cullen_2021} for the Keck/MOSFIRE observations.

        \begin{figure*}
            \centering
            \includegraphics[width=\linewidth]{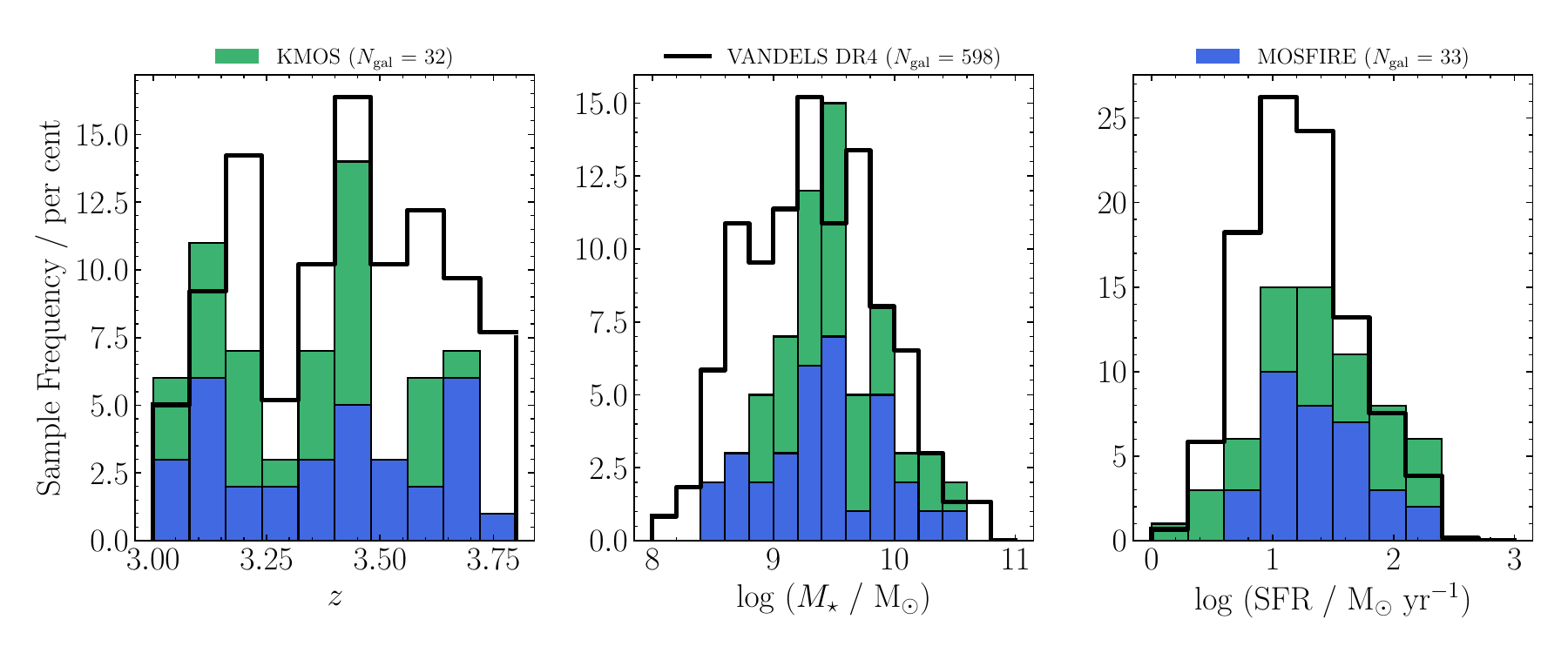}
            \caption{Plots showing, from left to right, the redshift, stellar mass, and star-formation rate distributions for the VLT/KMOS (green) and Keck/MOSFIRE (blue) samples. 
            The combined sample has an average redshift of $z \simeq 3.5$, an average stellar mass of $\mathrm{log}(M_{\star}/\mathrm{M}_{\odot}) = 9.5 \pm 0.5$ and an average SFR of $\mathrm{log(SFR/M}_{\odot}\mathrm{yr}^{-1}) = 1.4 \pm 0.5$.
            In all three panels, it can be seen that the distributions for both samples are comparable.
            Moreover, these distributions are consistent with the entire VANDELS DR4 $3.0 < z < 3.8$ star-forming sample shown in black, though the median SFR of our sample is marginally higher than the median SFR of the VANDELS sample ($\mathrm{log(SFR/M}_{\odot}\mathrm{yr}^{-1}) = 1.2 \pm 0.4$).
            The comparison highlights the fact that our sample is consistent with typical VANDELS star-forming galaxies and therefore representative of normal, main-sequence, star-forming galaxies at $z\simeq3.5$ \citep{McLure_2018}.}
            \label{fig:sample-histograms}
        \end{figure*}

        \subsubsection{Rest-frame optical emission line fluxes and redshifts}

            Rest-frame emission line fluxes were measured by fitting Gaussian profiles to the 1D science spectra extracted from the final datacubes. 
            For each individual line in a given galaxy spectrum, we isolated the expected position of the feature using the systemic redshift, within which we applied a Bayesian model fitting algorithm to fit a Gaussian profile characterised by an amplitude, width, and line centroid. 
            All of the fitted line profiles were visually inspected, and the best-fitting fluxes, FWHMs and centroids were determined by taking the median of the resulting posterior distributions, with uncertainty bounds taken as the $16^{\text{th}}$ and $84^{\text{th}}$ percentiles. 
            The \oii \ emission line doublet was blended at the resolution of our observations, and so for these lines two Gaussian profiles of equal widths were fitted simultaneously. 
            An example fit to the galaxy KVS-208 is shown in Fig. \ref{fig:example-kmos}.
            Within our $1$ arcsec apertures, our data exhibit a median $2\sigma$ line flux sensitivity of $\simeq 2\times10^{-18}\, \rm{erg\,s^{-1}\,cm^{-2}}\,${\AA}$^{-1}$.

            \par
            
            For one galaxy in our sample (KVS-248), the observed emission lines exhibited broad wings, and thus were better described by fitting a combination of a broad and narrow Gaussian profiles centred at the same wavelength, but with differing velocity widths. 
            The FUV VANDELS spectrum of this galaxy contained no high ionization emission lines indicating AGN activity, and as such the broad features are attributable to outflows, and the narrow features to the nebular emission of the galaxy.
            For this galaxy, we used only the narrow nebular feature flux measurements in our analysis.

        \subsubsection{Stellar masses and star-formation rates}
        \label{sec:data-mass-sfr}

            To estimate stellar masses and star-formation rates, we utilised the multiwavelength photometry from the VANDELS photometric catalogues \citep{McLure_2018}. 
            At the typical redshift of the VLT/KMOS sample ($z \simeq 3.5$), the emission line fluxes of the \oii, \hbeta \ and \oiii \ features will contaminate the broadband photometry in the $H$- and $K$- bands \citep{Cullen_2021}. 
            As the VANDELS photometry was derived using $2$ arcsec diameter apertures, we first scaled our $1$ arcsec KMOS line fluxes using the average ratio of $2$ arcsec to $1$ arcsec flux derived from the reference star data\footnote{We note that galaxies are not point sources and will exhibit different scaling factors when using variable aperture sizes. 
            To verify our chosen approach, we measured \oiiinwl \ line fluxes from our galaxies in $1$ arcsec and $2$ arcsec apertures. 
            Reassuringly, we find an equivalent average scaling factor from this approach.}.
            As discussed in the following, we find that this average correction yields excellent consistency between SED-derived SFRs and SFRs estimated from the dust-corrected \hbeta \ line flux.
            For each galaxy, we constructed a model emission-line spectrum based on our line fits, which were then integrated over the appropriate filter profiles.
            The VANDELS photometry was then adjusted by subtracting these resulting integrated fluxes. 
            In the $H$-band, the corrections ranged from $-0.23 < \Delta H /\mathrm{mag} < 0.0$ with a median correction of $\Delta H = -0.06 \ \mathrm{mag}$, while corrections in the $K$-band ranged from $-1.63 < \Delta K /\mathrm{mag} < 0.0$ with a median correction of $\Delta K = -0.21 \ \mathrm{mag}$.
            These corrections are consistent with the emission line corrections found for the Keck/MOSFIRE sample described in \citet{Cullen_2021} who report median corrections of $\Delta H = -0.07 \ \mathrm{mag}$ and $\Delta K = -0.35 \ \mathrm{mag}$.

            \par
            
            The emission line-corrected photometry was modelled using \href{https://github.com/cschreib/fastpp}{\textsc{FAST++}} \citep{Schreiber_2018}, a modified SED fitting code based on the original \textsc{FAST} software \citep{Kriek_2009}. 
            Following \citet{Cullen_2021}, we used the \citet{Conroy_2009} flexible stellar population synthesis models, assuming solar metallicities\footnote{While we will show in our subsequent analysis (Section~\ref{sec:results}) that our galaxies exhibit O abundances of $0.2 - 0.7 \, \mathrm{Z}_\odot$ and Fe abundances of $0.1 - 0.3 \, \mathrm{Z}_\odot$, fitting the broadband photometry with solar metallicity models is required for consistency with the \citet{Sanders_2021} $\text{E(B-V)}_\text{neb}$ calibration (Section~\ref{sec:dust-correction}; Equation~\ref{eq:sanders-dust}) which we use to dust correct our nebular emission line flues in the absence of a direct Balmer decrement measurements.}, a \citet{Chabrier_2003} initial mass function (IMF), constant star-formation histories and the \citet{Calzetti_2000} dust attenuation curve.
            These SED-fitting parameters were chosen to ensure full consistency between this work and \citet{Cullen_2021}, as well as with the stellar-to-nebular dust correction calibration derived by \citet{Sanders_2021} (see Section~\ref{sec:dust-correction} below).
            The final fits yielded an estimate of galaxy stellar mass (\mstar), star-formation rate (SFR$_{\mathrm{SED}}$) and a model of the underlying stellar continuum. 
            Using the best-fitting stellar continuum models we corrected our measured \hbeta \ flux values for the effect of underlying stellar absorption.
            This correction resulted in a median increase of 2.8\% to our \hbeta \ flux measurements.

        \subsubsection{Dust-correcting nebular emission line fluxes}
            \label{sec:dust-correction}

            Accurate measurements of the gas-phase oxygen abundance require correcting the observed line fluxes for nebular dust attenuation.
            Conventionally, these corrections are made directly using measurements of the Balmer decrement (\halpha/\hbeta).
            However, our observations lacked spectral coverage of the \halpha \ feature.
            As an alternative, we used the calibration between stellar reddening ($\text{E(B-V)}_\text{stellar}$), $\mathrm{SFR_{SED}}$, redshift and nebular reddening (($\text{E(B-V)}_\text{neb}$)) derived by \citet{Sanders_2021}:
            \begin{equation} \label{eq:sanders-dust} 
            \begin{aligned}
                \text{E(B-V)}_\text{neb} = & \quad \text{E(B-V)}_\text{stellar} - 0.604 \\ 
                & \quad + 0.538 \times [\log(\text{SFR}_\text{SED}) - 0.20 \times (z-2.3)].
            \end{aligned}
            \end{equation}
            This calibration was derived from observations of galaxies at $z \simeq 2.3$ using measurements of $\text{E(B-V)}_\text{neb}$ from the Balmer decrement and $\text{E(B-V)}_\text{stellar}$ from SED fitting (using the same SED fitting assumptions described above).
            The resulting calibration provides an unbiased estimate of $\text{E(B-V)}_\text{neb}$ with an intrinsic scatter of $0.23$ magnitudes. 
            In 2/32 galaxies, both \hgamma \ and \hbeta \ were detected allowing $\text{E(B-V)}_\text{neb}$ to be measured from the \hgamma/\hbeta \ ratio. The measured $\text{E(B-V)}_\text{neb}$ are consistent with the $\text{E(B-V)}_\text{neb}$ derived from Equation~\ref{eq:sanders-dust}.
            Based on the estimated $\text{E(B-V)}_\text{neb}$ values from Equation~\ref{eq:sanders-dust} we corrected our observed emission line fluxes assuming a \citet{Cardelli_1989} extinction curve\footnote{The choice of this extinction law was motivated by evidence of the nebular attenuation of high-redshift star-forming galaxies closely following the \citet{Cardelli_1989} extinction curve \citep[e.g.,][]{Reddy_2020}.}.

            \par

            The self-consistency of our nebular and stellar dust corrections was verified by deriving star formation rates from the dust-corrected \hbeta \ line fluxes and comparing to $\mathrm{SFR_{SED}}$.
            We converted the dust-corrected \hbeta \ fluxes (scaled to $2$ arcsec diameter apertures) into estimates of the intrinsic \halpha \ line flux assuming a ratio of \halpha/\hbeta= 2.86, and derived star formation rates using the \citet{Hao_2011} \halpha-\text{SFR} conversion modified for a \citet{Chabrier_2003} IMF. 
            The \hbeta-derived SFRs are compared with $\mathrm{SFR_{SED}}$ in Fig.~\ref{fig:hb-sfr-vs-phot-sfr} and show excellent agreement.
            The average offset between the two estimates is $0.06 \pm 0.09$ dex, where the error in the offset is derived from the median absolute deviation ($\sigma = 1.4826 \times \rm{MAD}$).
            Reassuringly, the SFRs derived from SED fitting and from \hbeta exhibit excellent consistency, highlighting the self-consistency of our stellar and nebular dust correction.
            Fig.~\ref{fig:hb-sfr-vs-phot-sfr} also demonstrates that the VLT/KMOS and Keck/MOSFIRE samples exhibit an indistinguishable trend, indicating that our analysis is consistent with that of \citet{Cullen_2021}.

            The final redshift, stellar mass and star-formation rate distributions of the complete sample are shown in Fig.~\ref{fig:sample-histograms}.

    \subsection{Composite Spectra}
    \label{sec:compspec}

        Of the $32$ galaxies in the VLT/KMOS sample, we were able to determine individual gas-phase metallicities for $21/32$ and stellar metallicities for $11/32$, with $8/32$ galaxies having determinations of both. (see Table~\ref{tab:bian.s99} and Section~\ref{sec:analysis} below).
        In order to fully leverage the entire sample, we therefore employed spectral stacking techniques.
        In Section \ref{sec:results}, we present the results of splitting the sample into two bins split at the median mass of $10^{9.53} \mathrm{M_\odot}$.
        To stack the spectra, we converted each spectrum into luminosity density units using its spectroscopic redshift and correct for nebular dust attenuation using its $\text{E(B-V)}_\text{neb}$ and the \citet{Cardelli_1989} extinction law. 
        Each spectrum was normalised using its \oiiia \ luminosity to prevent biasing the composites in favour of the brightest objects, and then re-sampled onto a common rest-frame wavelength grid with a spectral sampling of $1${\AA} per pixel. 
        The final stacked spectrum was generated by applying a $3\sigma$ clipping and taking the median value at each wavelength, and the associated error spectrum was calculated via bootstrap resampling.

        \par
        
        For the Keck/MOSFIRE galaxies, we adopt the sample selection and stack sizes described in \citet{Cullen_2021}. 
        As described in \citet{Cullen_2021}, $5/33$ galaxies were removed from the stacking sample due to the lack of spectral coverage of the \hbeta \ line, and two stacks were generated from the remaining 28 galaxies split at the median mass of $10^{9.4} \mathrm{M_\odot}$. 
        The stacks were generated using an identical methodology to the VLT/KMOS composites.

        \par

        As the emission features in the composite spectra were not necessarily well described by Gaussian profiles, we measured the integrated line fluxes via direct integration after subtracting any local continuum. 
        To infer uncertainties on line fluxes, we perturbed the composite spectrum by its associated error spectrum $500$ times and re-measured the line fluxes.
        For each line, the resulting flux and uncertainty were taken as the median and $68$th percentile width of the resulting distribution of measurements.
        To correct for stellar absorption in the \hbeta \ line, we multiplied the measured \hbeta \ line fluxes by the median correction factor for the individual galaxies ($\simeq \times 1.028$). 
        The properties of the composites are summarised in Table~\ref{tab:composite-spectra}.

        \par

        For the generation of the VANDELS stacks we followed a similar procedure except that in this case we normalised each individual spectrum by the median flux within the wavelength window $1420 \leq \lambda_{\rm{rest}} \leq 1480$.
        We also do not explicitly correct for dust attenuation when constructing the VANDELS stacks as the attenuation is accounted for in the stellar metallicity fitting methodology described in Section~\ref{sec:determine_zstar}.
        Each individual spectrum was resampled onto a common rest-frame wavelength grid between $1200 - 2000${\,\AA} with a spectral sampling of $1${\AA}.
        The final stacked spectrum was generated by applying a $3\sigma$ clipping and taking the median value at each wavelength, and the associated error spectrum was calculated via bootstrap resampling.

        \subfile{../tables/table3.tex}

%% file: tables/table1.tex
\begin{table*}
\caption{Summary of the VLT/KMOS observations.}
\label{tab:observations}
\renewcommand{\arraystretch}{1.3}
\begin{tabularx}{\textwidth}{@{\extracolsep{\fill}}ccccccc@{}}
\toprule
Pointing & Dates Observed & RA      & DEC     & $\mathrm{N_{gal}}$ & Exposure Time & Median FWHM Seeing\\ 
         &                & (J2000) & (J2000) &                    & (mins)        & (arcsec)     \\ \midrule \midrule
p1 & Oct $-$ Dec 2021                             & 03:32:54.9 & -27:44:42.2 & 20 & 566.6 & 0.56 \\
p2 & Dec 2021 $-$ Feb 2022 / Sep 2022 $-$ Feb 2023  & 03:32:47.2 & -27:54:45.7 & 20 & 266.6 & 0.57 \\ \bottomrule
\end{tabularx}
\end{table*}

%% file: tables/table2.tex
\begin{table}
\caption{Redshift, mass, gas-phase and stellar metallicity measurements for the KMOS sample utilising the \citetalias{Bian_2018} calibration scheme and the S99 stellar population models. For each galaxy, we also provide the corresponding VANDELS ID.}
\label{tab:bian.s99}
\renewcommand{\arraystretch}{1.3}
\setlength{\tabcolsep}{1pt}
\begin{tabularx}{\linewidth}{@{\extracolsep{\fill}}cccrcc@{}}
\toprule
Name & ID &$z_{\mathrm{spec}}$ & $\log $($M_\star$/$\mathrm{M}_\odot$) & $\log $($Z_\mathrm{g}$/$\mathrm{Z}_\odot$) & $\log $($Z_\star$/$\mathrm{Z}_\odot$) \\ \midrule \midrule
KVS-006 & 002584 & $3.367$ & $9.84$ & \textemdash & $-0.30^{\, +0.07}_{-0.05}$ \\
KVS-009 & 237255 & $3.472$ & $9.60$ & \textemdash & \textemdash \\
KVS-014 & 001280 & $3.082$ & $9.68$ & \textemdash & \textemdash \\
KVS-055 & 105702 & $3.085$ & $10.26$ & $-0.06^{\,+0.05}_{-0.06}$ & $-0.67^{\,+0.06}_{-0.06}$ \\
KVS-067 & 235637 & $3.082$ & $8.97$ & $-0.18^{\,+0.07}_{-0.08}$ & \textemdash \\
KVS-070 & 027049 & $3.250$ & $9.04$ & \textemdash & \textemdash \\
KVS-075 & 012786 & $3.348$ & $9.36$ & $-0.31^{\,+0.08}_{-0.08}$ & \textemdash \\
KVS-082 & 230755 & $3.185$ & $8.90$ & $-0.23^{\,+0.08}_{-0.08}$ & \textemdash \\
KVS-085 & 101998 & $3.193$ & $9.83$ & $-0.32^{\,+0.07}_{-0.09}$ & $-0.93^{\,+0.09}_{-0.07}$ \\
KVS-087 & 005891 & $3.081$ & $9.35$ & \textemdash & $-0.64^{\,+0.06}_{-0.06}$ \\
KVS-093 & 016368 & $3.423$ & $9.34$ & \textemdash & \textemdash \\
KVS-100 & 003496 & $3.031$ & $9.35$ & $-0.47^{\,+0.08}_{-0.08}$ & \textemdash \\
KVS-101 & 027379 & $3.591$ & $9.54$ & \textemdash & \textemdash \\
KVS-131 & 014729 & $3.607$ & $10.03$ & $-0.46^{\,+0.08}_{-0.09}$ & \textemdash \\
KVS-141 & 002312 & $3.470$ & $9.63$ & $-0.31^{\,+0.09}_{-0.09}$ & \textemdash \\
KVS-150 & 103010 & $3.462$ & $9.57$ & $-0.39^{\,+0.08}_{-0.09}$ & \textemdash \\
KVS-156 & 019593 & $3.426$ & $9.58$ & \textemdash & $-0.71^{\,+0.08}_{-0.08}$ \\
KVS-202 & 130840 & $3.182$ & $10.26$ & $-0.01^{\,+0.05}_{-0.06}$ & \textemdash \\
KVS-204 & 132586 & $3.472$ & $9.17$ & $-0.42^{\,+0.09}_{-0.10}$ & $-0.61^{\,+0.06}_{-0.06}$ \\
KVS-208 & 228618 & $3.188$ & $9.51$ & $-0.37^{\,+0.07}_{-0.08}$ & $-0.98^{\,+0.03}_{-0.03}$ \\
KVS-215 & 207213 & $3.386$ & $9.20$ & \textemdash & \textemdash \\
KVS-220 & 232419 & $3.424$ & $10.44$ & \textemdash & \textemdash \\
KVS-227 & 017345 & $3.611$ & $9.10$ & $-0.43^{\,+0.07}_{-0.08}$ & $-1.06^{\,+0.04}_{-0.04}$ \\
KVS-248 & 231194 & $3.078$ & $9.53$ & $-0.25^{\,+0.06}_{-0.06}$ & $-0.61^{\,+0.07}_{-0.06}$ \\
KVS-266 & 208053 & $3.089$ & $8.97$ & \textemdash & \textemdash \\
KVS-298 & 106745 & $3.465$ & $9.46$ & $-0.41^{\,+0.08}_{-0.09}$ & \textemdash \\
KVS-312 & 101861 & $3.239$ & $9.53$ & $-0.41^{\,+0.08}_{-0.08}$ & $-0.83^{\,+0.04}_{-0.03}$ \\
KVS-340 & 232407 & $3.411$ & $9.27$ & $-0.42^{\,+0.07}_{-0.08}$ & \textemdash \\
KVS-361 & 018915 & $3.605$ & $9.98$ & $-0.31^{\,+0.08}_{-0.09}$ & \textemdash \\
KVS-391 & 131717 & $3.071$ & $9.68$ & $-0.23^{\,+0.06}_{-0.06}$ & $-0.94^{\,+0.04}_{-0.05}$ \\
KVS-414 & 004151 & $3.651$ & $9.11$ & $-0.52^{\,+0.12}_{-0.13}$ & \textemdash \\
KVS-423 & 234108 & $3.341$ & $9.55$ & $-0.50^{\,+0.07}_{-0.08}$ & \textemdash \\ \bottomrule
\end{tabularx}
\end{table}

%% file: tables/table3.tex
\begin{table*}
\caption{Properties of the low-$M_\star$ and high-$M_\star$ composite spectra from the VLT/KMOS sample, utilising the \citetalias{Bian_2018} calibrations and S99 SPS models. We have excluded 
\neiiinwl/\oiinwl \ due to neither composite yielding a significant detection of \neiii. The same information for the low- and high-$M_\star$ composite spectra from the Keck/MOSFIRE sample are given in table 1 of \citet{Cullen_2021}.}

\renewcommand{\arraystretch}{1.3}
\label{tab:composite-spectra}
\begin{tabularx}{\textwidth}{@{\extracolsep{\fill}}rrcccccc@{}}
\toprule
Stack & Mass Range & \textbf{Median} $\log $($M_\star$/$\mathrm{M}_\odot$) & \textbf{log}(\oiiinwl/\hbeta) & \textbf{log}(\oiiinwl/\oiinwl) & \textbf{log}((\oiiinwl + \oiinwl)/\hbeta) & $\log $($Z_\mathrm{g}$/$\mathrm{Z}_\odot$) & $\log $($Z_\star$/$\mathrm{Z}_\odot$) \\ \midrule \midrule
Low-$M_\star$ & 8.9 $-$ 9.53 & $9.23$ & $0.78\pm0.15$ & $0.54\pm0.09$ & $1.02\pm0.15$ & $-0.46\pm0.05$ & $-0.95\pm0.05$ \\
High-$M_\star$ & 9.53 $-$ 10.44 & $9.68$ & $0.57\pm0.09$ & $0.36\pm0.07$ & $0.85\pm0.08$ & $-0.32\pm0.04$ & $-0.77\pm0.08$ \\ \bottomrule
\end{tabularx}
\end{table*}

%% file: sections/3-Analysis.tex
    The primary focus of this study is the determination of the gas-phase metallicities (\zgas, tracing O/H) and stellar metallicities (\zstar, tracing Fe/H) as a function of galaxy stellar mass for our sample.
    For the remainder of the paper we will primarily use the notation \zgas \ and \zstar \ when referring to the respective metallicity estimates.
    In the following section, we describe in detail how each of these properties was measured from the spectroscopic data.

    \subsection{Determination of gas-phase metallicity}
    \label{sec:determine_zgas}

        We estimated \zgas \ using ratios of the strong nebular emission lines measured from the rest-frame optical spectra, employing a variety of empirical strong-line calibrations.
        As discussed above, our $H$- and $K$-band spectroscopic observations cover the \oii, \neiii, \hbeta \ and \oiii \ emission lines, which permit the use of the following line ratio diagnostics:
        \begin{enumerate}
            \item $\mathrm{O32} =$ \oiii  \ / \oii
            \item $\mathrm{O3} =$ \oiiia \ / \hbeta
            \item $\mathrm{O2} =$ \oii \ / \hbeta
            \item $\mathrm{R23} =$ (\oiii \ + \oii) / \hbeta
            \item $\mathrm{Ne3O2} =$ \neiii \ / \oii
        \end{enumerate}
        To estimate \zgas \ from these ratios, we follow the methodology described in \citet{Cullen_2021} and use the empirical calibrations of \citet{Bian_2018} (hereafter \citetalias{Bian_2018}) as our primary calibration set.
        These calibrations are based on local analogues of high-redshift galaxies.
        We have additionally implemented the calibrations of \citet{Curti_2020} (hereafter \citetalias{Curti_2020}) to which we added the new $\mathrm{Ne3O2}$ line ratio calibration from \citet{Curti_2023}, derived from galaxies from SDSS. 
        Finally, we also implemented the calibration scheme of  \citet{Sanders_2023} (hereafter \citetalias{Sanders_2023}), which are derived from early \emph{JWST} measurements of auroral lines to determine direct O/H measurements of galaxies from $2 \leq z \leq 9$. 
        As \citet{Cullen_2021} only report \citetalias{Bian_2018}-based oxygen abundances for the Keck/MOSFIRE sample, we have calculated O/H for the Keck/MOSFIRE sample using these additional calibration schemes.
        The oxygen abundances corresponding to these alternative calibration schemes are reported in Table~\ref{tab:alternative-abundances}.
        
        \par

        For a given calibration scheme, we calculated the following $\chi^2$ equation: 
        \begin{equation} \label{eq:chi2} 
            \chi^2(x) = \sum_i \frac{(\text{R}_{\text{obs},i} - \text{R}_{\text{cal}, i}(x))^2}{(\sigma^{2}_{\text{obs},i} + \sigma^{2}_{\text{cal},i})},
        \end{equation}
        where the sum over $i$ represents the above set line ratios, ${x = 12 + \log(\text{O/H})}$\footnote{${x = 12 + \log(\text{O/H}) - 8.0}$ for the \citetalias{Sanders_2023} calibration.}, $\mathrm{R}_{\text{obs},i}$ is the logarithm of the $i$th line ratio, R$_{\text{cal},i}$ predicted value of $\mathrm{R}_{\text{obs},i}$ at $x$, and $\sigma_{\text{cal},i}$ and $\sigma_{\text{obs},i}$ are the uncertainties on the calibration and measured line ratio respectively. 
        We converted Equation~\ref{eq:chi2} into a likelihood function ($\propto e^{-\chi^2/2}$) which was then maximised over $x$ using the Bayesian nested sampling package \textsc{dynesty} \citep{dynesty_Speagle} assuming a uniform prior across the range $6.7 \leq 12 + \log(\mathrm{O/H}) \leq 9.0$ ($\simeq 0.01 - 2Z_\odot$).
        The final metallicity was taken to be the median of the resulting metallicity posterior distribution, with the 1$\sigma$ uncertainty derived from the $68$th percentile width of the posterior probability distribution.
        The resulting best-fitting $12 + \log(\mathrm{O/H})$ was converted to a gas-phase metallicity using $12 \ + \ \log(\text{O/H}) = \log(\text{Z}_\text{g}/\text{Z}_{\odot}) + 8.69$.         
        
        We note that our assumed prior extends beyond the formal range of the \citetalias{Bian_2018}, \citetalias{Curti_2020} and \citetalias{Sanders_2023} calibrations, which are ${7.8 \leq 12 + \log(\text{O/H}) \leq 8.4}$, ${7.6 \leq 12 + \log(\text{O/H}) \leq 8.9}$ and ${7.4 \leq 12 + \log(\text{O/H}) \leq 8.3}$ respectively. 
        These extrapolations were necessary because for some objects, the observed line ratios fell outside of the range of the calibrations.
        For the \citetalias{Bian_2018} and \citetalias{Curti_2020} calibrations, the number of such cases were low.
        For the \citetalias{Bian_2018} calibration, 8/41 individual galaxies fell into the extrapolated metallicity range, however, all of the composites estimates fell within the original range.
        For \citetalias{Curti_2020} calibration, all of the individual galaxies and composite fell within the defined metallicity range.
        The \citetalias{Sanders_2023} calibrations yielded the largest number of best-fit metallicities beyond the formal range (17/41), however all of the composites were consistent with the original range within $\simeq 2\sigma$.
        Because our main results are based primarily on the composite spectra, these extrapolations do not affect our main conclusions. 
        
        \par

        For a given object, not all emission lines were significantly detected ($> 3 \sigma$).
        We determined that the minimum requirement for a robust metallicity estimate was significant detections of the \oiiia \ and \oii \ features, since the O$_{32}$ ratio is needed to break the degeneracy of the double-valued O$_3$ calibration. 
        Ne$_3$O$_2$ is also a monotonic ratio and thus could be used alone to infer metallicity \citep{Shapley_2017}, but due to the weaker line strength of \neiii \ we found that there were no instances in which this was the only available line ratio.

        \par

        For our VLT/KMOS sample, $11/32$ galaxies were fit using all line ratios, $9/32$ galaxies were fit using all excluding $\mathrm{Ne3O2}$, and $1/32$ galaxies were fit using only $\mathrm{O_{32}}$; $11/32$ galaxies do not have individual metallicity determinations due to the lack of a significant estimate of the $\mathrm{O_{32}}$ ratio.
        For both the low- and high-\mstar \ composite spectra, we found that all lines except for \neiiinwl \ were significantly detected. 
        Consequently, the composite spectra did not have accurate estimates of $\mathrm{Ne3O2}$, and were fit using all other available ratios. 
        The resulting gas-phase metallicity measurements for the individual and composite spectra can be found in Tables~\ref{tab:bian.s99} and~\ref{tab:composite-spectra} respectively.
        Incorporating the estimates from the \citet{Cullen_2021} Keck/MOSFIRE sample adds an additional $21$ individual galaxies and $2$ composite spectra with \zgas \ estimates. 
        In total, therefore, our sample contains $42$ galaxies with individual \zgas \ estimates and $4$ \zgas \ estimates from composite spectra binned by stellar mass.

    \subsection{Determination of stellar metallicity}
    \label{sec:determine_zstar}

        We estimated \zstar \ from the VANDELS rest-frame FUV spectra following the methodology described in \citet{Cullen_2019} and \citet{Cullen_2021}. 
        Within the wavelength range $\lambda = 1221 - 2000${\,\AA}, SPS models were fit to the full spectrum, masking out regions defined by \citet{Steidel_2016} as containing nebular emission and ISM emission/absorption features. 

        \par

        As our fiducial model for the FUV spectral fitting we use the Starburst99 (S99) \citep{Leitherer_1999, Leitherer_2010}. 
        We assumed constant star-formation over timescales of 100 \Myr\footnote{We note that this choice of star-formation timescale is somewhat arbitrary as the FUV SEDs reach an equilibrium after a few $\times 10^7$ years (see the discussion in \citealp{Topping_2020B}).}, and generated models adopting the weaker-wind Geneva tracks with single-star evolution and no stellar rotation. 
        Five models were generated with metallicities of \zstar \ $= (0.001, 0.002, 0.008, 0.014, 0.040)$, which were then convolved to match the dispersion of the VANDELS spectra ($2.5$\AA) and resampled onto a grid of $1${\,\AA} per pixel. 
        In order to properly populate the \zstar \ parameter space, we constructed models of intermediate metallicities by interpolating between models of known metallicity. 
        To constrain the overall continuum shape of the observed spectra, we additionally fit for dust attenuation using the FUV attenuation curve parameterisation of \citet{Salim_2018} (following \citealp{Noll_2009}), which utilises three parameters $A_V$, $\delta$ and $B$ that characterise the normalisation, slope and UV bump strength of the dust attenuation curve respectively. 
        Full details of the model fitting and examples are given in \citet{Cullen_2019} and \citet{Cullen_2021}.

        \par

        In addition to our fiducial model we also constructed S99 models including the effect of stellar rotation in order to assess the level of systematic uncertainty arising from SPS model choices.
        The iron abundances corresponding to the models including stellar rotation can be found in Table~\ref{tab:alternative-abundances}.
        We chose not to employ the Binary Population and Stellar Synthesis (BPASS) models \citep[v2.2.1][]{Stanway_Eldridge_2018}, though \citet{Cullen_2019} explored the effect of substituting the S99 models for BPASS, finding that iron abundances were reduced by $\sim0.1$ dex. 
        Although the newer subset of BPASS models \citep{Byrne_2022} include $\alpha$ enhancement, it is not accounted for in the spectra of O/B type stars \citep{Chartab_2023}. 
        For our purposes of fitting the FUV, these newer models are essentially unchanged from BPASS v2.2.1.

        \par

        We used the \textsc{python} package \textsc{dynesty} to fit the spectral models and dust attenuation profile to the FUV spectra. 
        We used a log-likelihood function of the form:

        \begin{equation}
        \begin{aligned}\label{eq:dynesty_likelihood}
            \ln (L) = K - \frac{1}{2}\sum_{i} \left[ \frac{(f_i - f(\theta)_i)^2}{\sigma^2_i} \right] \
                   = K - \frac{1}{2} \chi^2,
        \end{aligned}
        \end{equation}
         where $K$ is a constant, $f$ is the observed flux, $f(\theta)$ is the model flux given a set of parameters $\theta$, and $\sigma$ is the error on the observed flux. 
         We sum the likelihood over all of the pixels in the spectra corresponding to the stellar continuum, using `mask 1' from \citet{Steidel_2016} to exclude wavelength pixels contaminated by ISM absorption and/or nebular emission lines. 
         For each parameter, the final value was taken as the $50$th percentile of the posterior probability distribution, and the 1$\sigma$ uncertainties were taken as the $68$ percentile width.
         Based on the results of \citet{Topping_2020B}, we do not report \zstar \ estimates for individual spectra with a signal-to-noise ratio of $< 5.6$ per resolution element which, as well as being poorly constrained, are also likely to return biased \zstar \ estimates.

        \par

        From the VLT/KMOS sample, $11/32$ galaxies yielded robust individual estimates of \zstar, as well as both of the low- and high-\mstar \ composite spectra. 
        The results of the individual and composite fitting for the VLT/KMOS galaxies can be found in Tables~\ref{tab:bian.s99} and~\ref{tab:composite-spectra} respectively. 
        For consistency, we also re-fit the individual and composite galaxy spectra from the \citet{Cullen_2021} Keck/MOSFIRE sample, yielding a further $6$ individual \zstar \ estimates.
        In total, therefore, our sample contains $17$ galaxies within individual \zstar \ estimates and $4$ \zstar \ estimates from composite spectra binned by stellar mass.

%% file: sections/4-MZRs.tex
\begin{figure}
            \centering
            \includegraphics[width=\linewidth]{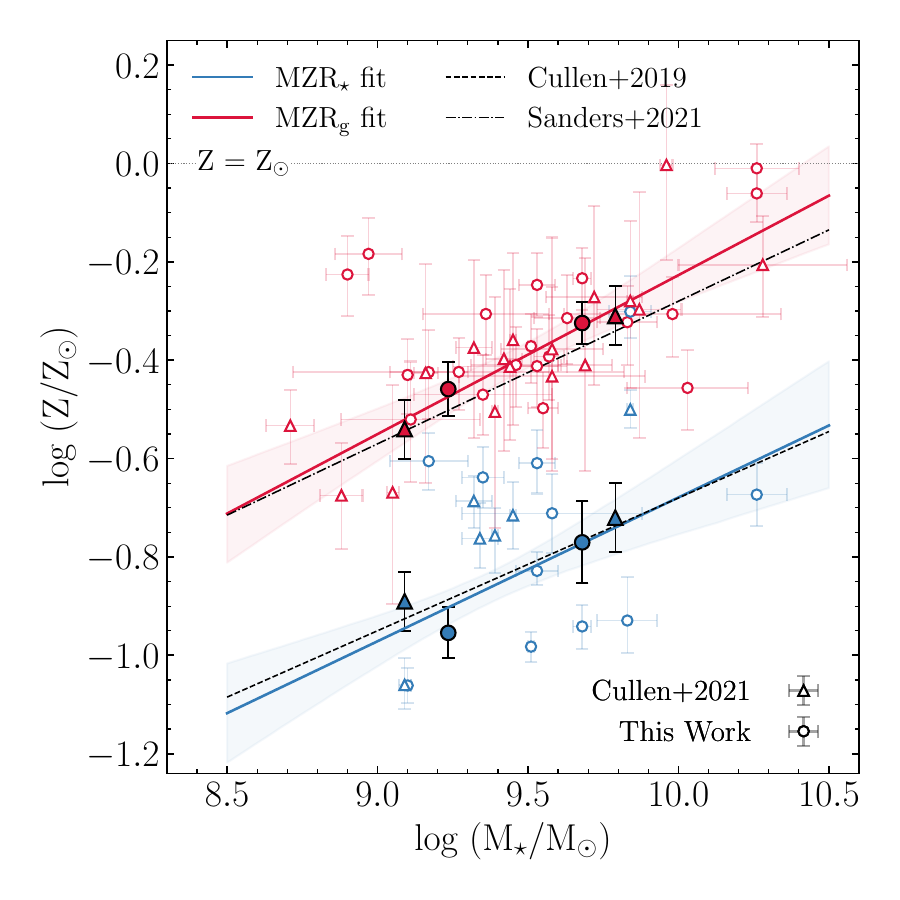}
            \caption{The gas-phase metallicities (red, tracing O/H) and stellar metallicities (blue, tracing Fe/H) for the VLT/KMOS (circles) and Keck/MOSFIRE galaxies \citep[triangles;][]{Cullen_2021} as a function of stellar mass.
            The individual galaxy determinations are shown by unfilled data points whilst the composite determinations are shown by the filled and outlined data points. 
            The best fitting gas-phase and stellar mass-metallicity relations (MZR$_\mathrm{g}$ and MZR$_\star$) are shown by red and blue solid lines respectively. $1\sigma$ uncertainties on each relationship are shown by shaded regions in their respective colours.
            For comparison, we plot literature relations at similar redshifts: the black dashed lines shows the MZR$_\mathrm{g}$ from \citet{Sanders_2021} and the black dot-dashed line shows the MZR$_\star$ from \citet{Cullen_2019}.
            The solar metallicity value (i.e., $\mathrm{log}(Z_{\star}/\mathrm{Z}_\odot) = 0.0$) is marked by the horizontal grey dashed line.} 
            \label{fig:combined-mzr}
    \end{figure}
\begin{figure*}
                \centering
                \includegraphics[width=\linewidth]{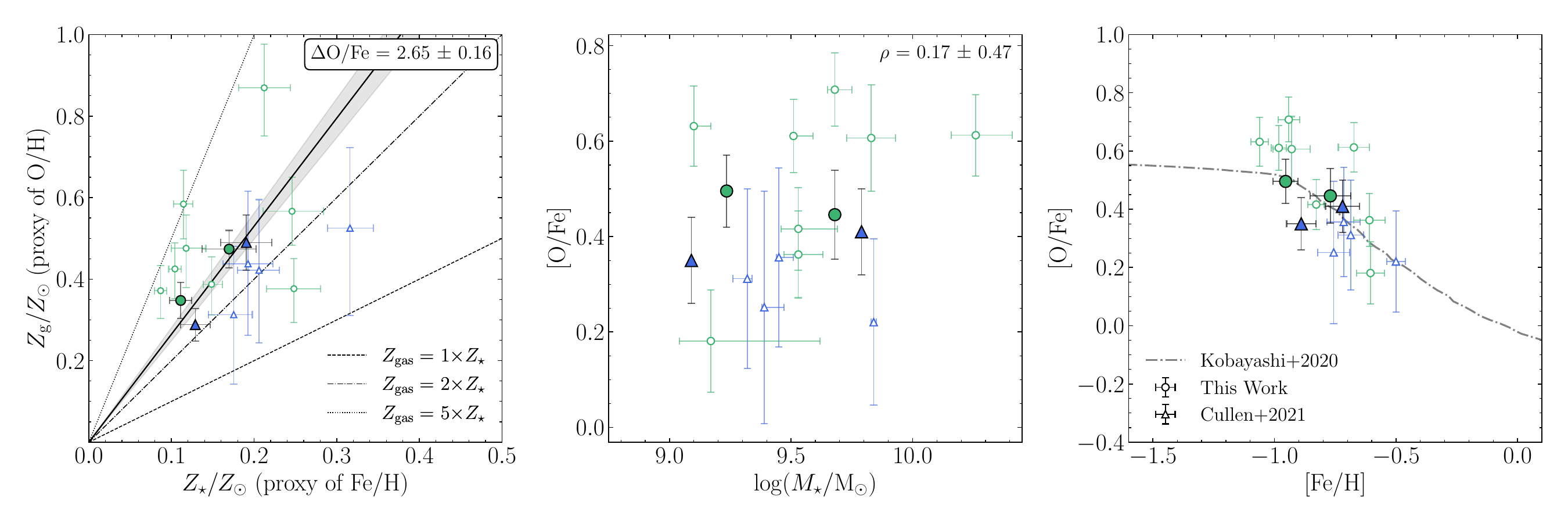}
                \caption{
                The O/Fe enhancement in our full $z\simeq3.5$ star-forming galaxy sample.
                In all panels, galaxies from our VLT/KMOS sample are shown as green circles, and galaxies from our Keck/MOSFIRE sample are shown as blue triangles. 
                The filled data points represent values derived from composite spectra, and the unfilled data points show individual galaxy determinations.
                The left-hand panel shows the $Z_{\mathrm{g}}/\mathrm{Z}_{\odot}$ (tracing O) versus $Z_{\star}/\mathrm{Z}_{\odot}$ (tracing Fe) relation for our sample. 
                The three black lines denote constant ratios from $1 \times$ to $5 \times$ the solar O/Fe value (see legend).
                Our sample lies consistently above the 1:1 line, consistent with an average enhancement of $\langle \mathrm{O/Fe} \rangle =2.65 \pm 0.16 \mathrm{(O/Fe)_\odot}$, as shown by the solid black line (with the grey shaded region highlighting the $68$ per cent confidence interval). 
                The centre panel shows the [O/Fe] versus \mstar \ relation for our sample.
                We find that [O/Fe] is not strongly stellar mass-dependent in the range $10^9 < M_{\star}/\mathrm{M}_{\odot} < 10^{10}$.
                In the right-hand panels we show the position of our sample in the [O/Fe] versus [Fe/H] diagram (see text for definitions) compared to the model evolutionary sequence for the Milky Way taken from \citet{Kobayashi_2020} (dot-dashed line).
                The galaxies in our sample overlap with the knee of the \citet{Kobayashi_2020} relation, indicating similarities between our high-redshift systems and the expected chemical evolution of Milky Way-like galaxies.
                }
                \label{fig:zgas-vs-zstar}
        \end{figure*}

    The aim of this work is to constrain the mass-metallicity scaling relations for O/H (tracing the gas phase, \zgas) and Fe/H (tracing massive stars, \zstar) of our sample, and to investigate the typical O/Fe ratio at $z\simeq3.5$. 
    In this section, we summarise our findings regarding both gas-phase and stellar mass-metallicity relationships (shown in Fig.~\ref{fig:combined-mzr}) and draw comparisons with relations from the literature (Fig.~\ref{fig:app-mzr-comparison}).
    Subsequently, for galaxies with simultaneous determinations of both O/H and Fe/H, we evaluate O/Fe in order to ascertain the degree of $\alpha$-enhancement present in our sample, as well as evidence for any dependence on stellar mass (Fig.~\ref{fig:zgas-vs-zstar}).

    \subsection{The gas-phase mass-metallicity relation}

        \begin{figure*}
            \centering
            \includegraphics[width=0.85\linewidth]{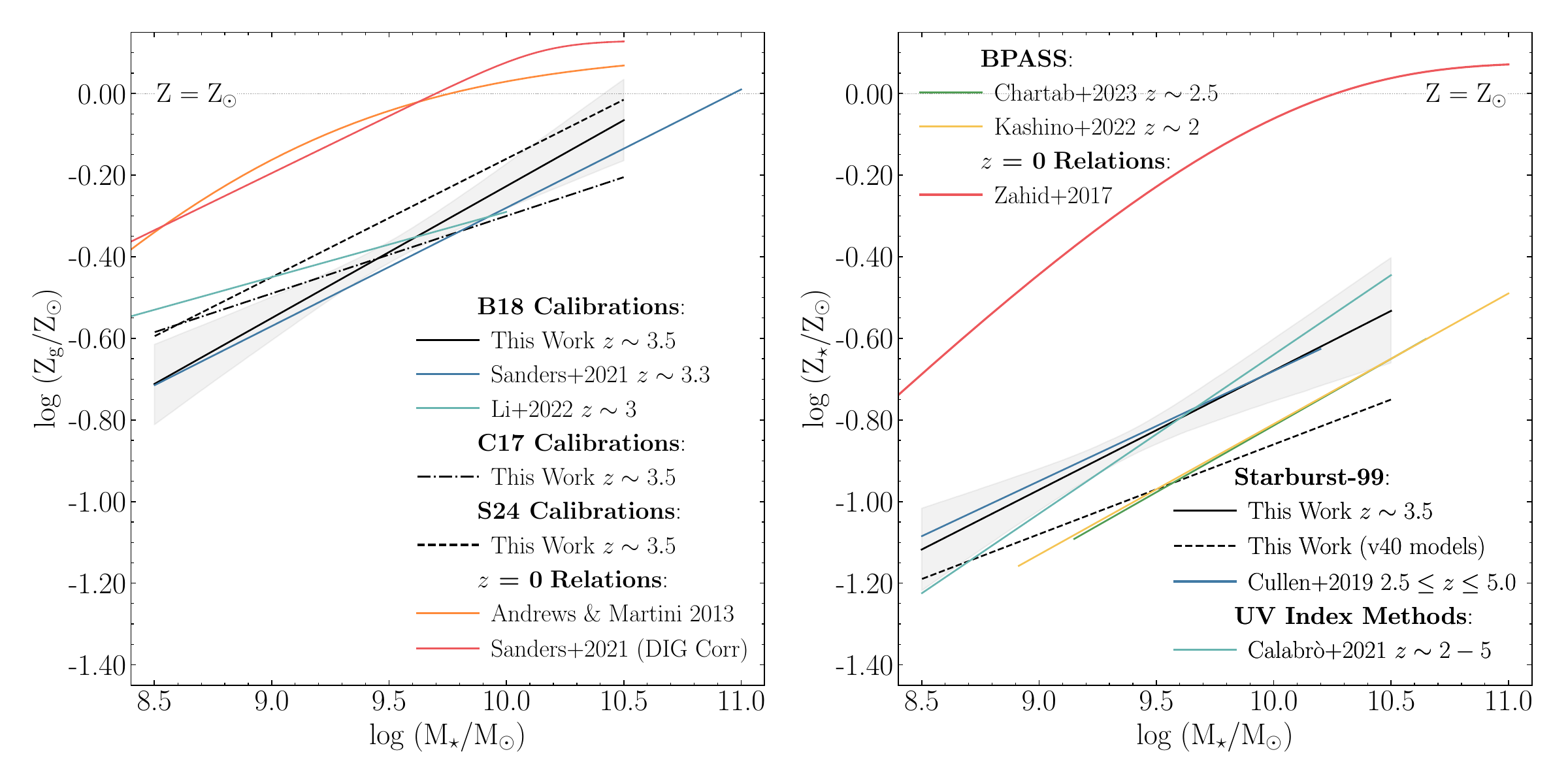}
            \caption{
            A comparison between the gas-phase (left panel) and stellar (right panel) mass-metallicity relationships of this work and other literature relations at $z=0$ and $z\simeq2-3$.
            Our fiducial relations (from Fig. \ref{fig:combined-mzr}) are plotted as solid black lines in each panel. 
            In the left panel we show the MZR$_\mathrm{g}$ determined from our sample using the \citetalias{Curti_2020} and \citetalias{Sanders_2023} calibrations as the black dot-dashed and black dashed lines, respectively.
            Various other literature relations are shown at $z=3.5$ and $z=0$, as indicated in the legend (see text for discussion).
            Despite systematic differences, the gas-phase O abundances at $z=3.5$ typically fall within the range $\simeq 20 - 70$ per cent solar across the stellar-mass range of our sample, a factor $\simeq 2 \times$ lower than local galaxies of the same stellar mass.
            In the right panel, we show our fiducial MZR$_\star$ along with the relation derived using S99 models including stellar rotation (black dashed line).
            We show various other literature relations at $z\simeq2-3$ and $z=0$, as indicated in the legend.
            Again, despite some differences, the high-redshift stellar Fe abundances typically fall within the range $\simeq 5 - 30$ per cent solar, in this case a factor $\simeq 4-5 \times$ lower than local galaxies of the same stellar mass.
            }
            \label{fig:app-mzr-comparison}
        \end{figure*}

        In Fig~\ref{fig:combined-mzr}, we show the gas-phase mass-metallicity relationship (hereafter MZR$_\mathrm{g}$) for our full sample assuming our fiducial strong line calibration scheme \citepalias{Bian_2018}. 
        It can be seen that for both individual galaxies and composite spectra, we find a clear trend between \zgas \ and \mstar, such that gas-phase O/H increases with increasing stellar mass.
        Fitting a linear relation in log-log space to the composite data yields a relationship of the form:
        \begin{equation}\label{eq:mzr_g}
            \log (Z_\mathrm{g}\mathrm{/Z_\odot}) = (0.32 \pm 0.09)m_{10} - (0.23 \pm 0.05),
        \end{equation}
        where $m_{10} = \log(M_\star/10^{10}\mathrm{M_\odot})$.
        This relation is in excellent agreement with the $z\simeq3.3$  MZR$_\mathrm{g}$ of \citet{Sanders_2021} and \citet{Cullen_2021}, which infer a slope and normalisation of $0.29 \pm 0.02$ and $-0.28 \pm 0.03$, respectively (also using the calibration of \citetalias{Bian_2018}).

        \par
        
        As can be seen from Fig.~\ref{fig:app-mzr-comparison}, the slope of the $z=3.5$ relation is consistent with the low mass slope (below $M_{\star} \simeq 10^{10} \mathrm{M}_{\odot}$) of the $z\sim0$ relations of \citet{Sanders_2021} and \citet{Andrews_2013}) (the latter based on a direct T$_{\mathrm{e}}$-based determination). 
        This comparison implies an approximately constant offset of $\simeq 0.3$ dex in $\log(\mathrm{Z_g})$ (i.e., a factor $2$ in O/H) between $z=0$ and $z=3.5$ in the stellar mass range $M_{\star} = 10^{8.5} - 10^{10.5} \mathrm{M_\odot}$.
        At $z=3.5$, we find that the oxygen abundance across this stellar mass range varies from $0.2 - 0.7\,\mathrm{Z_\odot}$, with an average value of $\simeq 0.4\,\mathrm{Z_\odot}$ at the median stellar mass of our sample ($M_{\star} \simeq 10^{9.5} \mathrm{M_\odot}$).
        At $z=0$, the typical O/H at the same stellar mass is $\simeq 0.8\,\mathrm{Z_\odot}$.
        Interestingly, these relations suggest that star-forming galaxies within the first $\simeq 15$ per cent of the Universe's history already contain roughly half the amount of oxygen present in the ISM of local $z=0$ star-forming galaxies with the same stellar mass.

        \par

        In Fig.~\ref{fig:app-mzr-comparison} we also show how the form of the MZR$_\mathrm{g}$ for our $z=3.5$ sample changes depending on the choice of strong-line calibration.
        For the \citetalias{Sanders_2023} calibration, we measure a slope of $0.29 \pm 0.13$ consistent with our fiducial model, but offset to higher metallicities with a normalisation of $-0.16 \pm 0.07$.
        For the \citetalias{Curti_2020} calibrations, we measure a similar normalisation to our fiducial model ($-0.30 \pm 0.03$) but a shallower slope of $0.19 \pm 0.07$.
        Clearly, subtle strong-line systematics are present, which we discuss further below.
        Nevertheless, the range of metallicities is fairly consistent, with O/H $\simeq40-50$ per cent solar of the solar value at $M_{\star} \simeq 10^{9.5} \mathrm{M_\odot}$.

        \par

        We note that below $M_{\star} \simeq 10^{9} \mathrm{M_\odot}$, we currently lack data, and the extrapolation of MZR$_\mathrm{g}$ given by Equation~\ref{eq:mzr_g} is naturally highly uncertain in this stellar mass regime.
        Indeed, recent \emph{JWST} studies, using similar strong-line approaches, find evidence for a flattening of the MZR$_\mathrm{g}$ slope at  $M_{\star} \lesssim 10^{9} \mathrm{M_\odot}$ \citep[e.g.,][]{Li_2023, He_2023}.
        For the purposes of this work we restrict ourselves to the simple linear form of the MZR$_\mathrm{g}$ (in log-log space), noting that our results are most robust in the stellar mass range $M_{\star} \simeq 10^{9}-10^{10} \mathrm{M_\odot}$.

    \subsection{The stellar mass-metallicity relation}

        In Fig~\ref{fig:combined-mzr}, we also show the stellar mass-metallicity relationship (hereafter MZR$_\star$) for our full sample assuming our fiducial S99 model.
        Despite fewer individual \zstar \ constraints, both the individual galaxies and the composites again show clear evidence for a stellar mass-metallicity relationship. 
        Fitting to the composite data yields a relationship of the form:
        \begin{equation} 
            \log (Z_{\star}\mathrm{/Z_\odot}) = (0.30 \pm 0.11)m_{10} - (0.68 \pm 0.07),
        \end{equation}
        where $m_{10} = \log(M_\star/10^{10}\mathrm{M_\odot})$. 
        This relation is in excellent agreement with the results presented \citet{Cullen_2019}, who measured a slope and normalisation of $0.27 \pm 0.06$ and $-0.68 \pm 0.04$ respectively based on fitting composite spectra generated from a much larger VANDELS sample ($\simeq 700$ individual galaxies) using S99 models (see also \citealp{Cullen_2021}).
        Across the stellar mass range of our sample, we find that stellar Fe abundances vary from $0.08 - 0.3\,\mathrm{Z_\odot}$, with a mean value of $\simeq 0.15\,\mathrm{Z_\odot}$ at $M_{\star} \simeq 10^{9.5} \mathrm{M_\odot}$.
        
        From Fig. \ref{fig:app-mzr-comparison} it can again be seen that for $M_{\star} \lesssim 10^{10} \mathrm{M}_{\odot}$ the MZR$_\star$ slope at $z=3.5$ is consistent with the local universe determination \citet{Zahid_2017}.
        Other local estimates (not shown) yield similar results \citep[e.g.,][]{Galazzi_2005}.
        As with the gas-phase relation, we see an offset in the normalisation of the MZR$_\star$, such that the galaxy distribution evolves towards higher metallicity at fixed mass at later times.
        However, the evolution of stellar Fe abundances is more pronounced than gas-phase O abundances, increasing by $\simeq 0.6$ dex (i.e., a factor of $4$) between $z=3.5$ and $z=0$ (cf. a factor $2$ for O/H).
        The reason for this is naturally explained by the fact that galaxies at $z=3.5$ are typically $\alpha$-enhanced (i.e., Fe-deficient), as we discuss in the following.

        \par
        
        We compare our results to other estimates of MZR$_\star$ at $z>2$ in Fig. \ref{fig:app-mzr-comparison}.
        We find good agreement between the slope of our relation at $M_{\star} \lesssim 10^{10} \mathrm{M}_{\odot}$ and the relations of \citet{Kashino_2022} at $z\simeq2.2$ (slope $=0.32 \pm 0.03$) and \citet{Chartab_2023} at $z\simeq2.5$ (slope $= 0.32 \pm 0.02$)\footnote{We note that \citet{Chartab_2023} do not provide a linear fit to their data, so this value represents our own fit to their published data.}. 
        Both \citet{Kashino_2022} and \citet{Chartab_2023} use a methodology similar to ours (i.e., full spectra fitting to rest-frame FUV spectra) but use the BPASS SPS models and analyse two independent datasets: zCOSMOS \citep{Lilly_2007} and LATIS \citep{Newman_2020} respectively. 
        We also find good agreement with the $2 < z < 5$ relation of \citet{Calabrò_2021}, who estimate stellar metallicities using a subset rest-frame UV absorption features for a sample of VANDELS galaxies.
        The consistency in the resulting MZR$_\star$ shape regardless of the stellar model choice and methodology is therefore reassuring.
        The normalisation of both $z \simeq 2.5$ relations is lower by $\simeq 0.12$ dex compared to our fiducial relation, but this can be reconciled by the fact that fitting with BPASS models is known to return a lower \zstar \ by roughly $0.1$ dex \citep[e.g.,][]{Cullen_2019}.
        We note that fitting with the S99 models including stellar rotation also yields lower estimates of \zstar \ and reduces the difference with the BPASS estimates (Fig. \ref{fig:app-mzr-comparison}).
        Nevertheless, across a variety of modelling assumptions and datasets, a consistent pictures emerges of galaxies at $z \simeq 2-3$, with $M_{\star} \lesssim 10^{9} - 10^{10} \mathrm{M}_{\odot}$ having stellar Fe abundances of $\simeq 5$ to $20$ per cent of the solar value.
        Crucially, the estimates \emph{all} fall below the MZR$_{\mathrm{g}}$ at the same redshift.

    \subsection{Enhanced O/Fe abundance ratios}
    \label{sec:alpha-enhancement}

        A comparison the MZR$_\mathrm{g}$ and MZR$_\star$ relations in Fig. \ref{fig:combined-mzr} shows that the gas-phase relation (tracing O/H) is offset by $\simeq + 0.4$ dex compared to the stellar relation (tracing Fe/H) across the full stellar mass range.
        This offset represents direct evidence of enhanced O/Fe ratios (relative to the solar value) in star-forming galaxies at $z=3.5$.
        Throughout the discussion below (and the remainder of the paper), we adopt the following definitions:
        \begin{equation}
        \mathrm{[Fe/H]} = \mathrm{log(Fe/H)-log(Fe/H)}_{\odot},
        \end{equation}
        (i.e., the Fe abundance relative to the solar value) and,
        \begin{equation}
        \mathrm{[O/Fe]} = \mathrm{[O/H]-[Fe/H]},
        \end{equation}
        (i.e., the O/Fe ratio relative to the solar value).
        In the context of this work, $\mathrm{log}(Z_{\star}/\mathrm{Z}_{\odot})$ is an estimate of [Fe/H] and $\mathrm{log}(Z_{\mathrm{g}}/Z_{\star})$ is an estimate of [O/Fe].
        A value of $\mathrm{[O/Fe]}>0$ represents an enhancement of O/Fe relative to the solar value (i.e., $\alpha$-enhancement).

        \par

        We find that the low- and high-mass composites from the VLT/KMOS sample have highly consistent [O/Fe] ratios of ${0.49 \pm 0.07}$ and ${0.45 \pm 0.09}$ respectively. 
        For the Keck/MOSFIRE stacks of \citet{Cullen_2021}, we measure [O/Fe] values of $0.35 \pm 0.09$ for the low-mass composite and $0.41 \pm 0.09$ for high-mass composite. 
        Each of these values is statistically significant at the $\gtrsim 4 \sigma$ level and also consistent with the inverse variance-weighted mean value $0.5 \pm 0.03$ of the 12/65 individual galaxies with robust O/H and Fe/H abundances in our sample.
        In the left panel of Fig.~\ref{fig:zgas-vs-zstar} we show a linear fit to our composites in the \zstar-\zgas \ plane, from which we derive a typical value of $\mathrm{O/Fe} = 2.65 \pm 0.16  \times (\mathrm{O/Fe})_\odot$ (i.e., $\mathrm{[O/Fe]}=0.42 \pm 0.03$).
        Our new constraint is in excellent agreement with the value of $\mathrm{O/Fe} = 2.57 \pm 0.38 \times (\mathrm{O/Fe})_\odot$ derived from the Keck/MOSFIRE sample in \citet{Cullen_2021}, and reduces the uncertainty on the degree of enhancement by a factor of $\sim 2$.
        Overall, we find that our sample is ubiquitously $\alpha$-enhanced.

        \par

        The consistent slopes of the MZR$_\mathrm{g}$ and MZR$_\star$ suggest no stellar-mass dependence, which we show explicitly in the middle panel of Fig~\ref{fig:zgas-vs-zstar}.
        In the stellar mass range $M_{\star} = 10^9 - 10^{10} \mathrm{M}_{\odot}$ we see no evidence for a deviation from the full-sample average.
        At face value, this result suggests that the same physical processes are responsible for setting the shape of the low mass slope of the stellar and gas-phase MZRs (namely large scale galactic outflows; e.g., \citealp{Cullen_2019}, \citealp{Sanders_2021}), and that the typical star-formation histories are similar for galaxies with $M_{\star} = 10^{9} \mathrm{M}_{\odot}$ and $M_{\star} = 10^{10} \mathrm{M}_{\odot}$.
        We discuss the physical interpretation of our results in more detail in Section \ref{sec:gce-modelling}.
        It is worth noting, however, that this conclusion is calibration-dependent. 
        Applying the strong-line calibration of \citet{Curti_2020} yields a shallower MZR$_\mathrm{g}$ (Fig. \ref{fig:app-mzr-comparison}) such that lower-mass galaxies are more $\alpha$-enhanced.
        In Fig. \ref{fig:zg-zs-comparison-s99} and Appendix \ref{app:systematic_uncertainties} we explore the extent to which the choice of strong-line calibration affects the derived value of O/Fe.
        Crucially, whilst the degree of $\alpha$-enhancement varies across the tested calibration schemes and SPS models, all cases yield [O/Fe] ratios enhanced by a factor of $\gtrsim 2.5$.
        
        \par

        In the right-hand panel of Fig. \ref{fig:zgas-vs-zstar} we show the location of our $z=3.5$ star-forming galaxies in the [O/Fe] versus [Fe/H] plane compared to the one-zone Galactic Chemical Evolution model of \citet{Kobayashi_2020}.
        The \citet{Kobayashi_2020} model is based on an assumed Milky Way-like star formation history and is shown to match a variety of stellar archaeological data in the Milky Way (MW).
        For our fiducial strong-line calibration and SPS model assumptions, our data provide a good match to the predicted evolution of MW-like galaxies.
        The $z=3.5$ galaxies fall close to the `knee' in the relation at $\mathrm{[Fe/H]} \simeq -1.0$ (i.e., the point at which Fe enrichment from SN Ia initiates the transition to lower O/Fe). 
        The consistency between our high-redshift systems and the trend of MW stars represented by the \citet{Kobayashi_2020} is encouraging.
        
        \par
        
        Our results are generally in good agreement with other works in the literature that have estimated O/Fe at $z>2$ \citep[e.g.,][]{Steidel_2016, Cullen_2019, Topping_2020A, Kashino_2022, Strom_2022, Chartab_2023}.
        Of the studies shown in Fig.~\ref{fig:app-mzr-comparison}, \citet{Kashino_2022} and \citet{Strom_2022} estimate $\mathrm{O/Fe} = 2.12 \pm 0.54 \times (\mathrm{O/Fe})_\odot$ and $\mathrm{O/Fe} \simeq 2.2 \times (\mathrm{O/Fe})_\odot$ respectively; \citet{Chartab_2023} measure an average enhancement of ${\mathrm{O/Fe} \simeq 3.7 \pm 0.4 \times (\mathrm{O/Fe})_\odot}$.
        \citet{Steidel_2016} and \citet{Topping_2020A} report larger values of up to $5-7 \times (\mathrm{O/Fe})_\odot$ but crucially all studies agree that star-forming galaxies at $z>2$ are commonly $\alpha$-enhanced (generally, studies utilising the BPASS SPS models suggest lower [Fe/H] and hence larger [O/Fe]; \citealp{Cullen_2021}).
        Combined, the results of all these studies suggest a picture in which the ISM of star-forming galaxies at $z>2$ is dominated by CCSNe enrichment.

%% file: sections/5-Outflows.tex
        Previous studies have demonstrated how the power-law slope of the MZR (both stellar and gas-phase) below $\mathrm{log}(M_{\star}) \simeq 10.5$ can be used to constrain the mass-dependence of galaxy-scale outflows \citep[e.g.,][]{Cullen_2019, Sanders_2021, Chartab_2023}.
        The strength of galaxy-scale outflows is also predicted to affect the degree of $\alpha$-enhancement in the ISM \citep[e.g.,][]{Weinberg_2017}.
        In this section, we employ a chemical evolution model to simultaneously interpret the scaling of O and Fe abundances with stellar mass as well as the typical O/Fe ratios in our sample.

        \subsection{An analytical chemical evolution model}

        We employ the analytical one-zone chemical evolution model of \citet{Weinberg_2017} (hereafter \citetalias{Weinberg_2017}) which incorporates a realistic delay time distribution (DTD) for Type-Ia SNe, allowing the evolution of iron-peak elements to be tracked separately to $\alpha$-elements.
        The models assume a constant (i.e., time-independent) star-formation efficiency ($\mathrm{SFE; SFE} = \dot{M}_\star / M_\mathrm{gas}$) in accordance with the `linear Schmidt law' and mass outflow rates are assumed to scale with the star-formation rate, with the scaling factor described by the mass-loading factor ($\eta = \dot{M}_{\mathrm{out}} / \dot{M}_{\mathrm{\star}}$). 
        Therefore, the accretion and loss of gas (i.e., the baryon cycle) as a function of time is defined by the star formation history (SFH), the star formation efficiency (which we characterise by the gas depletion timescale $t_\mathrm{dep} = 1 / \mathrm{SFE}$), and the mass loading factor $\eta$. 
        We adopt a linearly rising SFH to best match the predicted SFH of star-forming galaxies \citep[e.g.,][]{Reddy_2012, Ciesla_2017}, and use the $t_\mathrm{dep}$-\mstar \ scaling relation of \citet{Tacconi_2018} to set the star formation efficiency.

        Unless otherwise stated, we adopt the fiducial model parameters from \citetalias{Weinberg_2017}.
        The Type-Ia DTD is exponentially declining with a minimum delay time of 0.15 \Gyr \ and an \textit{e}-folding timescale of $1.5$\Gyr.
        We assume a \citet{Kroupa_2001} IMF, and the CCSNe yields of \citet{Chieffi_2004} and \citet{Limongi_2006} in which $1.5\mathrm{M_\odot}$ of O and $0.12 \mathrm{M_\odot}$ of Fe are returned to the ISM for every $100 \mathrm{M_\odot}$ of star formation.
        The corresponding return of Fe from Type-Ia SNe is 0.17 $\mathrm{M_\odot}$. 
        The model assumes that gas accretion from the IGM is pristine, that outflowing gas is of equal metallicity to the metallicity of the star-forming ISM, and that enriched material from CCSNe are instantaneously returned and mixed into the star-forming ISM.
        The resulting time evolution of the abundances of O and Fe is given by equations $55-58$ in \citetalias{Weinberg_2017}.
        
        Our aim is to explore whether, using this model, we can find a simple scaling between $\eta$ and \mstar \ that can simultaneously explain both the stellar and gas-phase MZRs and the O/Fe ratios of our sample.
        Both theory and observations support a scenario in which lower-mass galaxies exhibit higher $\eta$ values compared to higher-mass galaxies (i.e., gas is more efficiently ejected in lower mass galaxies; \citealp{Hayward_2017, Llerena_2023}). 
        Ejection is also more likely to occur at velocities greater than their respective escape velocities, because momentum sources such as supernovae are more effective in lower mass galaxies \citep{Chisholm_2017}.
        We employ the simple scaling $\mathbf{\eta=\alpha (M_{\star}/10^{10}\mathrm{M_\odot})^{\beta}}$ suggested by the FIRE simulations \citep{Muratov_2015}. 
        This simple parameterisation has been shown to effectively reproduce the power-law slope of the MZR \citep[e.g.,][]{Cullen_2019}.

        \begin{figure*}
            \centering
            \includegraphics[width=\linewidth]{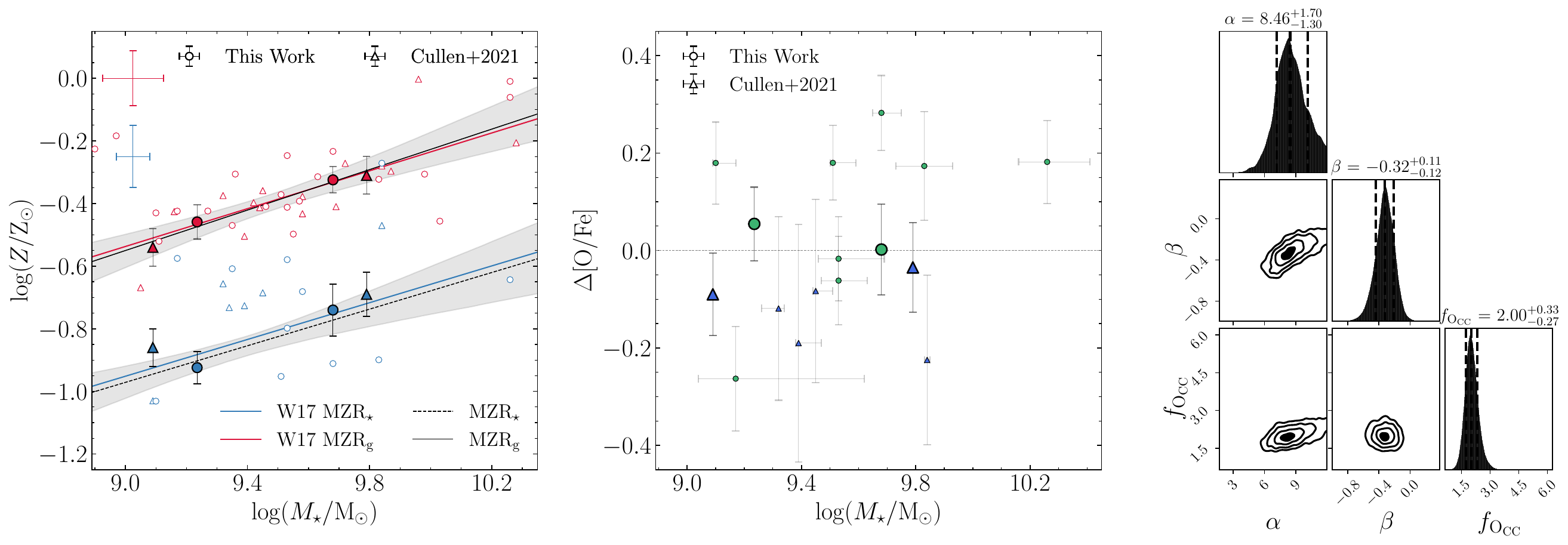}
            \caption{The results of chemical evolution model fitting to the O and Fe abundances in our sample, showing our constraints on the stellar-mass scaling of galaxy-scale outflows.
            Our principal result is that our data are well described by a model (following the framework of \citetalias{Weinberg_2017}) in which the mass-loading parameter $\eta$ ($= \dot{M}_{\mathrm{out}} / \dot{M}_{\mathrm{\star}}$) scales with stellar mass as $\eta \propto M_{\star}^{\beta}$ where $\beta = -0.32 \pm 0.12$.
            In the left-hand and centre panels, we show a comparison between the best-fitting model and our data. 
            In each panel the VLT/KMOS galaxies are shown as circular data points, and Keck/MOSFIRE galaxies as triangular data points.
            Composite galaxies are shown as large filled points with error bars, while individual galaxies are small open points.
            The median uncertainty for all individual galaxies is indicated in each panel.
            In the left-hand panel, we show our data and the resulting gas-phase and stellar mass-metallicity relations at $z \simeq 3.5$ for our best-fitting model as the red and blue solid lines, respectively.
            The resulting relations are in excellent agreement with the data.
            In the centre panel, we show the difference between our measured [O/Fe] and the predicted [O/Fe] from our model as a function of galaxy stellar mass. 
            Our best fitting model accurately predicts the [O/Fe]-enhancement of our sample and supports the stellar mass independence of $\alpha$-enhancement.
            The right-hand panel shows a corner plot giving the posterior probability distributions for the three parameters in our chemical evolution model fitting.
            }
            \label{fig:gce-parameter-fitting}
        \end{figure*}

        We also found that it was necessary to increase the CCSNe oxygen yields to match the data.
        The default \citetalias{Weinberg_2017} yields result in a maximum of $\mathrm{[O/Fe]} \simeq 0.4$ which falls below a number of our estimates (as well as other [O/Fe] estimates at high-redshift; e.g., \citealp{{Steidel_2016}}; \citealp{Chartab_2023}).
        Evidence from stellar data suggests that $\mathrm{[O/Fe]}$ can reach values up to $\simeq 0.6$ (e.g., APOGEE DR17 abundances; \citealp{Abdurro'uf_2022}).
        Alternative stellar yield estimates are also consistent with a $\mathrm{[O/Fe]} \simeq 0.6$ plateau \citep[e.g.,][]{nomoto2006, kobayashi2006, Kobayashi_2020}.
        Rather than fixing our yields using an alternative prescription, we include a free parameter in our model which scales the \citetalias{Weinberg_2017} CCSNe O yield, denoted $f_{\mathrm{O_{CC}}}$.
        We note that the need to scale the oxygen yields is in part a consequence of the underlying simplifying assumptions of the model, namely the assumption that inflowing gas is pristine and that outflows are not more enriched than the ISM. 
        For example, including enriched inflows from previous cycles of star formation would increase the degree of $\alpha$-enhancement, as the composition of the reaccreted gas in the inflows would reflect an earlier ISM enriched primarily by CCSNe.
        These enriched inflows might mitigate the need to enhance the CCSNe O yields.
        However, it is unlikely that the effect of these assumptions will change the mass scaling of the outflows, which is driven primarily by the mass scaling of the individual abundances.
        
        \subsection{Constraints on the $\eta$-\mstar \ relation}

        For a given composite galaxy and set of model parameters ($\alpha$, $\beta$ and $f_{\mathrm{O_{CC}}}$), we use our model to predict both the O and Fe abundances at the average redshift (i.e., time) and \mstar.
        We sampled the entire parameter space and derive the best-fitting values of the parameters with \textsc{dynesty}, employing a Gaussian likelihood (Equation \ref{eq:dynesty_likelihood}).
        We assumed a flat prior on each parameter: $0 < \alpha < 12$; $-1.4 < \beta < 0.4$ and $0.5 < \mathrm{f_{O_{CC}}} < 4$.
        The best-fitting parameters are shown in the right-hand panel of Fig.~\ref{fig:gce-parameter-fitting}.

        The consistency between the best-fitting model and data is highlighted in the left-hand and centre panels of Fig.~\ref{fig:gce-parameter-fitting}.
        In the left panel, we have converted the model into $M_{\star}-\mathrm{O/H}$ and $M_{\star}-\mathrm{Fe/H}$ relations at the median redshift of the sample ($z=3.5$). 
        It can be seen that both the slope and normalisation of both relations are reproduced well.
        Similarly, our model predicts minimal differences from the measured [O/Fe] across all of our individual galaxies and composites.
        We show this comparison in the centre panel as a function of stellar mass, again highlighting the excellent agreement.
        Overall, we find that this relatively simple chemical evolution model, which fundamentally assumes (i) linearly rising star formation histories, (ii) the \citet{Tacconi_2018} $t_\mathrm{dep}-M_{\star}$ relation, and (iii) a power-law anticorrelation between $\eta$ and \mstar, does a remarkably good job of reproducing the O- and Fe-abundance scaling relations and the flat [O/Fe] ratios of our sample.
        
\begin{figure}
            \centering
            \includegraphics[width=\linewidth]{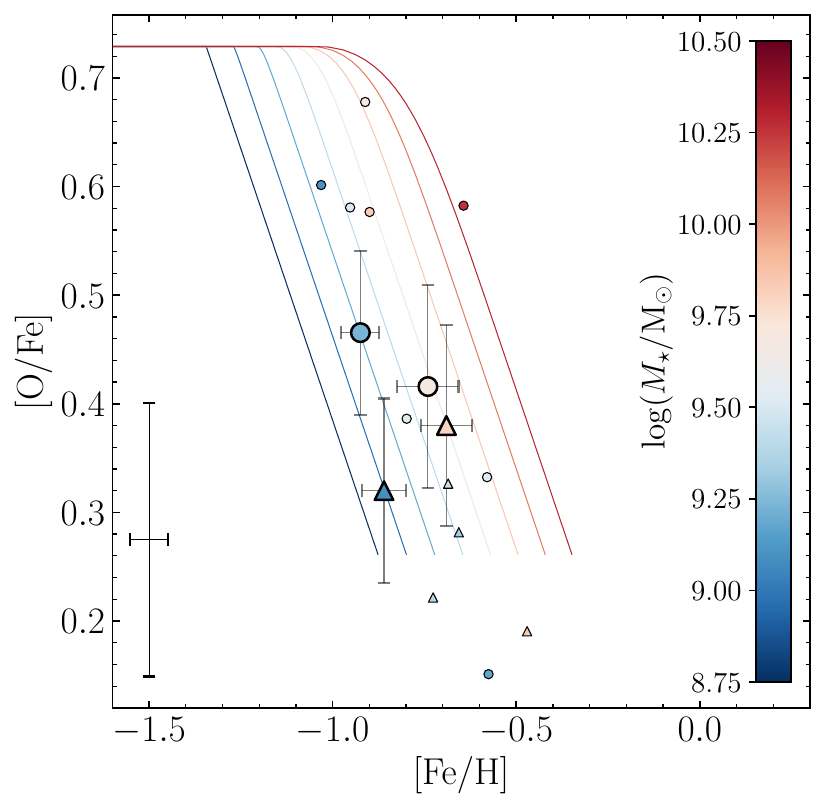}
            \caption{The resulting tracks in the [O/Fe] versus [Fe/H] plane derived from our chemical evolution model fitting to the O and Fe abundances in our sample.
            VLT/KMOS galaxies are shown as circular data points and Keck/MOSFIRE galaxies as triangular points.
            Composite galaxies are shown as large filled points with error bars while individual galaxies are smaller open points, with their average errors given by the error bar in the bottom left of the plot.
            Galaxies are colour-coded according to their stellar mass, shown by the inset colourbar on the right. 
            The coloured lines show the mass-dependent chemical evolution tracks for our best-fitting model parameters.
            These tracks are also temporal sequences, meaning that it is possible to trace the chemical abundances of each element as a function of time from the onset of star formation.
            In general, all data points fall close to the track corresponding to their stellar mass.}
            \label{fig:gce-tracks}
        \end{figure}

        In Fig.~\ref{fig:gce-tracks} we show the predicted time evolution of O and Fe abundances as a function of stellar mass for the best-fitting model, where some of the fundamental physical trends are apparent.
        First, galaxies with lower \mstar \ have lower [Fe/H]; in the context of our modelling, this is due to more efficient removal of enriched content at lower \mstar \ (i.e., larger $\eta$), and higher gas fractions diluting the ISM at lower \mstar \ \citep{Tacconi_2018}.
        Second, [O/Fe] is approximately constant at all [Fe/H], which is a consequence of the shallow $t_\mathrm{dep}$-\mstar \ relation and implies similar star formation timescales for galaxies in the stellar mass range $10^9 - 10^{10} \mathrm{M}_{\odot}$.
        We find a best-fitting oxygen yield enhancement of $f_{\mathrm{O_{CC}}} = 2.00^{+0.33}_{-0.27}$, which results in an [O/Fe] plateau of $\mathrm{[O/Fe]} = 0.73 \pm 0.06$.
        This value is somewhat higher than the $\mathrm{[O/Fe]} \simeq 0.6$ plateau of the \citet{Kobayashi_2020} models, but is broadly consistent within the uncertainties and probably acceptable given the significant uncertainties in supernova yields \citep[e.g.,][]{Romano_2010}.

\begin{figure}
            \centering
            \includegraphics[width=\linewidth]{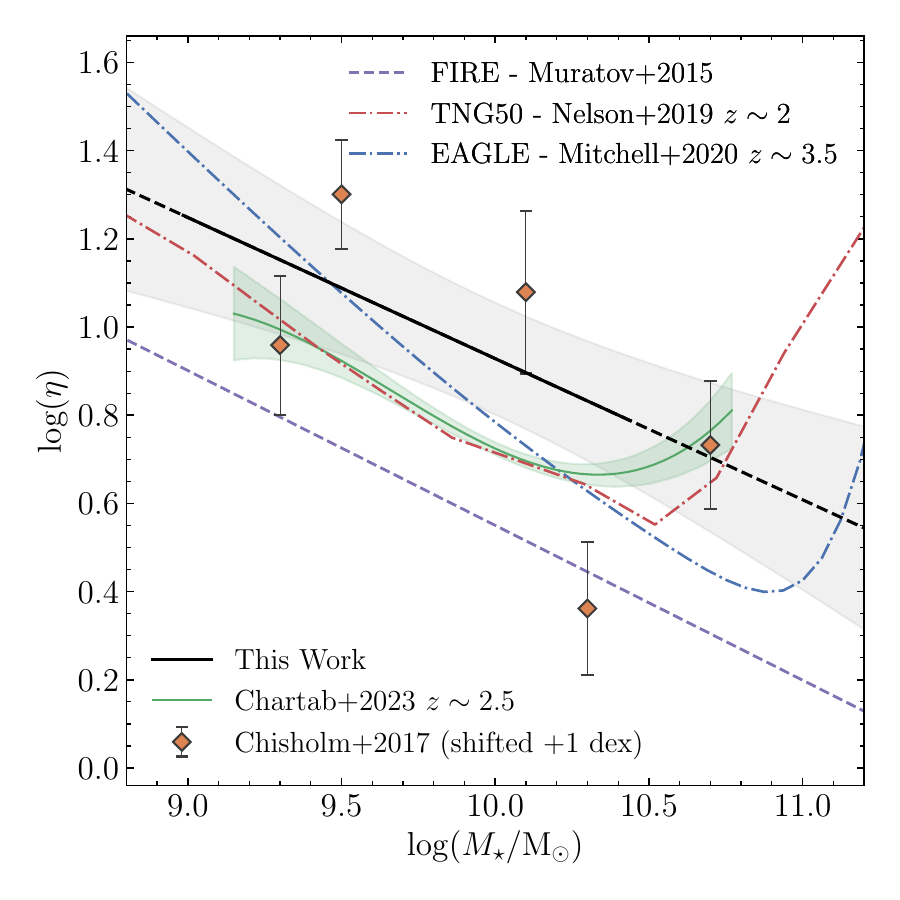}
            \caption{The mass loading factor ($\eta$) versus stellar mass (\mstar) relation derived from fitting to the O and Fe abundances in our sample within the chemical evolution modelling framework of \citet{Weinberg_2017} (black line).
            The dashed black line shows the extrapolation of our fitted relation outside the mass regime of our sample.
            The green curve shows the best-fit relation from \citet{Chartab_2023}, derived by fitting to their stellar MZR relation at $z \simeq 2.5$ (also using the \citealp{Weinberg_2017} analytic models). 
            We also show the $\eta-M_{\star}$ relations derived from the FIRE (\citealp{Muratov_2015}; dashed purple), EAGLE (\citealp{Muratov_2015}; dot-dashed blue) and TNG50 (\citealp{Nelson_2019}; dot-red blue) cosmological simulations.
            The orange diamond data points show the mass loading factors and stellar masses of four local star-forming galaxies taken from \citet{Chisholm_2017}.
            The \citet{Chisholm_2017} estimates for $\eta$ were derived using ionised FUV absorption lines, and we find that the absolute values need to be scaled by a factor of $\simeq 10$ to match our relation (see text for discussion).
            In general, both the observational data and the simulation predictions show a consistent $\eta-M_{\star}$ scaling, with $-0.4 \lesssim \beta \lesssim -0.3$ (where $\eta \propto M_{\star}^{\beta}$).
            Differences in the overall normalisation are relatively minor considering the various systematic uncertainties involved (e.g., yields, abundance estimates, the radius at which $\eta$ is defined in simulation analyses).}
            \label{fig:mass-loading-factor}
        \end{figure}

        In Fig.~\ref{fig:mass-loading-factor} we show our best-fitting relation between $\eta$ and $M_{\star}$ in comparison to predictions from various cosmological simulations and the recent constraint of \citet{Chartab_2023}.
        The \citet{Chartab_2023} constraint was derived by applying the \citetalias{Weinberg_2017} modelling framework to their $\mathrm{Fe/H}-M_{\star}$ relation derived from the LATIS survey.
        In the mass regime to which we are sensitive ($M_{\star} \lesssim 10^{10.3} \mathrm{M}_{\odot}$), the shape of the relations are in excellent agreement.
        \citet{Chartab_2023} find $\eta \propto M_{\star}^{-0.35 \pm 0.01}$ compared to our constraint of $\eta \propto M_{\star}^{-0.32 \pm 0.12}$.
        Unfortunately, our data do not probe the mass regime in which \citet{Chartab_2023} find evidence for an upturn in $\eta$ which they attribute to AGN-driven winds (at $M_{\star} \gtrsim 10^{10.5} \mathrm{M}_{\odot}$).
        We find systematically larger values of $\eta$ at fixed \mstar \ by $0.1-0.2$ dex but note that the absolute normalisation of the relation is strongly dependent on the measured O and Fe abundances and subject to larger systematic uncertainties.
        The mass dependence of $\eta$ derived here is also consistent with other similar models in the literature; for example, \citet{Cullen_2019} and \citet{Sanders_2021}, which find $\beta \simeq -0.4$ and $\beta = -0.35 \pm 0.01$, respectively. 

        In Fig. \ref{fig:mass-loading-factor} we also show data from \citet{Chisholm_2017} who estimated $\eta$ for a sample of local star-forming galaxies via ionised FUV ISM absorption lines.
        \citet{Chisholm_2017} find a mass scaling of $\eta \propto M_{\star}^{-0.4}$ but with a normalisation lower by a factor of ${\simeq 10}$.
        In Fig. \ref{fig:mass-loading-factor} we have scaled their $\eta$ estimates by this factor to compare with our relation and highlight the consistent mass dependence.
        The discrepancy in normalisation is not surprising, since the \citet{Chisholm_2017} relation traces only the ionised gas phase.
        Multiple studies have found that the ionised phase accounts for only a small fraction of the total outflowing gas, which is probably dominated by the molecular phase \citep[e.g.,][]{Fluetsch_2021, roberts-borsani_2020, Concas_2022, Llerena_2023, Weldon_2024}. 
        Scaling $\eta$ derived from the ionised gas by a factor $\sim 10-100$ to estimate the total mass loading factor is not out of the question.
        
        We compare with the predictions of the TNG50 \citep{Nelson_2019}, EAGLE \citep{Mitchell_2020} and FIRE \citep{Muratov_2015} cosmological simulations in Fig. \ref{fig:mass-loading-factor}.
        Generally, the agreement is encouraging.
        In particular, the mass scaling derived from the simulations is in good agreement with our derivation for $M_{\star} \lesssim 10^{10.5} \mathrm{M}_{\odot}$.
        Above this mass threshold, simulations that include AGN feedback (TNG-50, EAGLE) predict an upturn on $\eta$, however, our data are not sensitive to this mass regime.
        Differences in normalisation are evident (e.g., compared to the FIRE predictions) but we note that the normalisation derived from simulations is dependent on the radius at which $\eta$ is defined.
        Coupled with the yield and abundance uncertainties that affect our derived relation, we do not consider these discrepancies significant.
        Crucially, the form of the $\eta-M_{\star}$ relationship derived from a relatively simple chemical evolution model fit to O and Fe abundances of galaxies at high redshift is fully consistent with the predictions of more detailed physical simulations.

%% file: sections/6-Conclusions.tex
    In this work be have performed an analysis of $65$ star-forming galaxies at $z=3.5$ with ground-based near-IR spectra tracing rest-frame optical emission lines and ground-based optical spectra tracing the rest-frame FUV continuum.
    For both individual galaxies and composite spectra, we have derived estimates of the oxygen abundances from the \oiii, \oii \ and \hbeta \ nebular emission lines and iron abundances from full spectral fitting to the rest-frame FUV continuum.
    Combining our abundance determinations with stellar masses inferred from photometric SED fitting, we have derived scaling relations for both oxygen and iron abundances with stellar mass.
    For the galaxies and composite spectra with both oxygen and iron abundance estimates, we have determined the O/Fe ratio and investigated its dependence on stellar mass.
    Finally, we have utilised one-zone chemical evolution models to interpret our results in terms of star-formation timescales and the mass-dependence of large-scale outflows. 
    The main results of our study can be summarised as follows:

    \begin{enumerate}
    
        \item Combining dust-corrected strong emission line ratios from our rest-frame optical spectra with the strong-line calibration scheme of \citet{Bian_2018}, we derive gas-phase metallicities (\zgas, a proxy of O/H) for our individual galaxies and composites. 
        Comparing the \zgas \ measurements of our sample with their stellar masses, we observe a clear gas-phase mass-metallicity (MZR$_{\mathrm{g}}$) relationship (Fig.~\ref{fig:combined-mzr}). 
        Our relationship exhibits a slope of d(log \zgas)/d(log \mstar) $= 0.32 \pm 0.09$, with \zgas \ increasing from $\simeq 0.2\mathrm{Z_\odot}$ at \mstar = $10^{8.5} \ \mathrm{M}_\odot$ to $\simeq 0.7\,\mathrm{Z_\odot}$ at \mstar = $10^{10.5} \ \mathrm{M}_\odot$.
        
        \item 
        From fitting to the rest-frame FUV continuum spectra we derive stellar metallicities (\zstar, a proxy for Fe/H) and observe a clear stellar mass-metallicity (MZR$_{\star}$) relationship in our data (Fig.~\ref{fig:combined-mzr}). 
        The slope of the MZR$_{\star}$ is consistent with the slope of the MZR$_{\mathrm{g}}$, but exhibits a constant offset to lower \zstar \ across the fitting range of stellar mass. 
        Across the same range of mass (\mstar = $10^{8.5 - 10.5} \ \mathrm{M}_\odot$), we observe that \zstar \ increases from $\simeq 0.08\,\mathrm{Z_\odot}$ to $\simeq 0.5\mathrm{Z_\odot}$.
        
        \item For the galaxies and composites for which we have determinations of both \zstar \ and \zgas we calculate the O/Fe ratio. 
        We find that our sample displays super-solar O/Fe ratios (i.e., $\alpha$-enhancement; Fig.~\ref{fig:zgas-vs-zstar}) and find no evidence for a dependence on stellar mass in the range $M_{\star} = 10^{9} - 10^{10} \ \mathrm{M}_\odot$.
        Fitting to our data yields an average enhancement of ${\mathrm{(O/Fe)} = 2.65 \pm 0.16 \ \times \mathrm{(O/Fe)_\odot}}$.
        Our new estimates add further support to a picture in which star-forming galaxies at $z>2$ are ubiquitously $\alpha$-enhanced.
        
        \item 
        Using the one-zone analytical chemical evolution models of \citet{Weinberg_2017} we fit for the relative O and Fe abundances of our sample in the stellar mass range $M_{\star} = 10^{9} - 10^{10} \ \mathrm{M}_\odot$.
        We find that a model in which the gas depletion scales with stellar mass following \citet{Tacconi_2018}, and the mass-loading of stellar winds scales as $\eta \propto M_{\star}^{-0.32\pm0.12}$ provides an excellent description of our data.
        In this modelling framework, it is an increasing efficiency of stellar winds at lower stellar mass that is driving the gas-phase and stellar mass-metallicity relations.
        The enhanced O/Fe ratios are a product of delayed Fe enrichment from Typa-Ia supernovae.
        Interestingly, this $\eta-M_{\star}$ scaling is in good agreement with the predictions of hydrodynamical simulations (e.g. \citealp{Muratov_2015, Nelson_2019}; see Fig. \ref{fig:mass-loading-factor}).
        
    \end{enumerate}

    The sample sizes with both \zstar \ and \zgas \ determinations remain small, and while we have addressed some of the systematics involved in our determinations, complete high-redshift strong-line calibration schemes and stellar population models remain active areas of research and development. 
    Future surveys, such as AURORA (PI Shapley, Sanders; GO1914) and EXCELS (PI Carnall, Cullen; GO3543) will supply additional data and aid the assessment of these systematic effects, with which we will be able to further assess the degree of $\alpha$-enhancement in high-redshift star-forming galaxies.

%% file: tables/app1.tex
\begin{table}
\caption{Gas-phase and stellar metallicity measurements for the composite and individual KMOS sample, utilising the \citetalias{Sanders_2023} and \citetalias{Curti_2020} calibration schemes and the S99 v40 rotational models.}
\label{tab:alternative-abundances}
\renewcommand{\arraystretch}{1.3}
\begin{tabularx}{\linewidth}{@{\extracolsep{\fill}}cccc@{}}
\toprule
Name & S24 $\log (Z_\mathrm{g}/\mathrm{Z}_\odot)$ & C20 $\log (Z_\mathrm{g}/\mathrm{Z}_\odot)$ & v40 $\log (Z_\star/\mathrm{Z}_\odot)$ \\ \midrule \midrule
KVS-006 & \textemdash & \textemdash & $-0.85^{\,+0.01}_{-0.01}$ \\
KVS-009 & \textemdash & \textemdash & \textemdash \\
KVS-014 & \textemdash & \textemdash & \textemdash \\
KVS-055 & $0.05^{\,+0.09}_{-0.09}$ & $-0.24^{\,+0.03}_{-0.03}$ & $-0.96^{\,+0.05}_{-0.05}$ \\
KVS-067 & $-0.09^{\,+0.11}_{-0.13}$ & $-0.29^{\,+0.04}_{-0.04}$ & \textemdash \\
KVS-070 & \textemdash & \textemdash & \textemdash \\
KVS-075 & $-0.20^{\,+0.13}_{-0.14}$ & $-0.37^{\,+0.05}_{-0.05}$ & \textemdash \\
KVS-082 & $-0.14^{\,+0.11}_{-0.12}$ & $-0.31^{\,+0.04}_{-0.05}$ & \textemdash \\
KVS-085 & $-0.27^{\,+0.11}_{-0.12}$ & $-0.40^{\,+0.05}_{-0.05}$ & $-0.99^{\,+0.05}_{-0.05}$ \\
KVS-087 & \textemdash & \textemdash & $-0.80^{\,+0.04}_{-0.03}$ \\
KVS-093 & \textemdash & \textemdash & \textemdash \\
KVS-100 & $-0.43^{\,+0.13}_{-0.15}$ & $-0.50^{\,+0.05}_{-0.04}$ & \textemdash \\
KVS-101 & \textemdash & \textemdash & \textemdash \\
KVS-131 & $-0.39^{\,+0.12}_{-0.11}$ & $-0.45^{\,+0.04}_{-0.05}$ & \textemdash \\
KVS-141 & $-0.16^{\,+0.12}_{-0.15}$ & $-0.41^{\,+0.06}_{-0.07}$ & \textemdash \\
KVS-150 & $-0.35^{\,+0.11}_{-0.12}$ & $-0.39^{\,+0.04}_{-0.04}$ & \textemdash \\
KVS-156 & \textemdash & \textemdash & $-0.79^{\,+0.06}_{-0.05}$ \\
KVS-202 & $0.24^{\,+0.04}_{-0.06}$ & $-0.19^{\,+0.03}_{-0.03}$ & $-0.77^{\,+0.03}_{-0.03}$ \\
KVS-204 & $-0.40^{\,+0.11}_{-0.12}$ & $-0.42^{\,+0.04}_{-0.04}$ & $-0.72^{\,+0.04}_{-0.05}$ \\
KVS-208 & $-0.32^{\,+0.10}_{-0.11}$ & $-0.42^{\,+0.03}_{-0.03}$ & $-1.01^{\,+0.03}_{-0.03}$ \\
KVS-215 & \textemdash & \textemdash & \textemdash \\
KVS-220 & \textemdash & \textemdash & \textemdash \\
KVS-227 & $-0.40^{\,+0.11}_{-0.12}$ & $-0.46^{\,+0.04}_{-0.04}$ & $-1.10^{\,+0.03}_{-0.03}$ \\
KVS-248 & $-0.18^{\,+0.09}_{-0.09}$ & $-0.34^{\,+0.03}_{-0.03}$ & $-0.79^{\,+0.06}_{-0.05}$ \\
KVS-266 & \textemdash & \textemdash & \textemdash \\
KVS-298 & $-0.31^{\,+0.12}_{-0.14}$ & $-0.47^{\,+0.04}_{-0.04}$ & \textemdash \\
KVS-312 & $-0.42^{\,+0.10}_{-0.09}$ & $-0.41^{\,+0.03}_{-0.03}$ & $-0.89^{\,+0.03}_{-0.03}$ \\
KVS-340 & $-0.38^{\,+0.10}_{-0.12}$ & $-0.44^{\,+0.04}_{-0.04}$ & \textemdash \\
KVS-361 & $-0.23^{\,+0.11}_{-0.14}$ & $-0.37^{\,+0.04}_{-0.04}$ & \textemdash \\
KVS-391 & $-0.16^{\,+0.08}_{-0.09}$ & $-0.33^{\,+0.03}_{-0.03}$ & $-1.03^{\,+0.04}_{-0.04}$ \\
KVS-414 & $-0.50^{\,+0.25}_{-0.26}$ & $-1.22^{\,+0.71}_{-0.52}$ & \textemdash \\
KVS-423 & $-0.45^{\,+0.11}_{-0.11}$ & $-0.49^{\,+0.04}_{-0.04}$ & \textemdash \\ \bottomrule
\multicolumn{4}{c}{KMOS Composites} \\
\midrule 
Low-$M_\star$  & $-0.41^{\,+0.08}_{-0.08}$ & $-0.46^{\,+0.04}_{-0.04}$ & $-1.06^{\,+0.05}_{-0.05}$ \\
High-$M_\star$ & $-0.28^{\,+0.06}_{-0.06}$ & $-0.37^{\,+0.03}_{-0.03}$ & $-0.95^{\,+0.06}_{-0.05}$ \\ \bottomrule
\end{tabularx}
\end{table}